\documentclass[%
 reprint,
 superscriptaddress,
 amsmath,amssymb,
 aps,
 prl,
floatfix,
nolongbibliography,
]{revtex4-2}

\usepackage{graphicx}
\usepackage{dcolumn}
\usepackage{bm}
\usepackage{hyperref}
\usepackage[dvipsnames]{xcolor}
\hypersetup{
    colorlinks = true,
    linkcolor = blue,
    anchorcolor = blue,
    citecolor = blue,
    filecolor = blue,
    urlcolor = black
}

\usepackage{multirow}

\begin{document}

\preprint{APS/123-QED}

\title{
Microstructural Pattern Formation during Far-from-Equilibrium Alloy Solidification
}

\author{Kaihua Ji}
\affiliation{%
Physics Department and Center for Interdisciplinary Research on Complex Systems, Northeastern University, Boston, Massachusetts 02115, USA
}%
\author{Elaheh Dorari}
\affiliation{%
Physics Department and Center for Interdisciplinary Research on Complex Systems, Northeastern University, Boston, Massachusetts 02115, USA
}%
\author{Amy J. Clarke}
\affiliation{%
Department of Metallurgical and Materials Engineering, Colorado School of Mines, Golden, Colorado, 80401, USA
}%
\author{Alain Karma}%
 \email[]{a.karma@northeastern.edu}
\affiliation{%
Physics Department and Center for Interdisciplinary Research on Complex Systems, Northeastern University, Boston, Massachusetts 02115, USA
}%

\date{\today}

\begin{abstract}

We introduce a new phase-field formulation of rapid alloy solidification that quantitatively incorporates nonequilibrium effects at the solid-liquid interface over a very wide range of interface velocities. Simulations identify a new dynamical instability of dendrite tip growth driven by solute trapping at velocities approaching the absolute stability limit. They also reproduce the formation of the widely observed banded microstructures, revealing how this instability triggers transitions between dendritic and microsegregation-free solidification. Predicted band spacings agree quantitatively with observations in rapidly solidified Al-Cu thin films.
\end{abstract}

\maketitle





The past two decades have witnessed major progress in modeling complex interface patterns that form during alloy solidification. A major contributor to this progress has been the advent of the phase-field (PF) method \cite{Boettinger2002Phase-fieldSolidification,Steinbach2009Phase-fieldScience,Karma2016,Kurz2020Progress2018,Tourret2022Phase-fieldChallenges}, which circumvents front tracking by making the solid-liquid interface spatially diffuse over some finite width $\sim W$, and the development of quantitative PF formulations \cite{Karma1998,Karma2001,Echebarria2004,Folch2005QuantitativeGrowth,Plapp2011UnifiedFunctional,Boussinot2014AchievingContrast} that have enabled simulations on experimentally relevant length and timescales with a computationally tractable choice of $W$ on the pattern scale \cite{Haxhimali2006OrientationEvolution,Dantzig2013DendriticComputations,Bergeon2013SpatiotemporalSolidification,Clarke2017,Song2018Thermal-fieldSolidification,Ghosh2018SimulationSuperalloys,Wang2020ModelingModel}. 

Morphological instability driving microstructural pattern formation occurs over an extremely wide range of solidification velocities $V$ spanning six orders of magnitude from $\mu$m/s to m/s, with different ranges of $V$ relevant for different solidification processes from conventional casting to metal additive manufacturing \cite{kurz1989fundamentals,Dantzig2016Solidification}. To date, however, PF formulations to quantitatively simulate alloy solidification patterns have been primarily developed and validated for slow $V$ \cite{Karma2001,Echebarria2004}, conditions under which the solid-liquid interface can be assumed to remain in local thermodynamic equilibrium. While there have been attempts to extend quantitative modeling to rapid solidification, existing PF formulations have been limited to a small departure from equilibrium \cite{Pinomaa2019QuantitativeSolidification}, or have only reproduced solute trapping in one-dimensional (1D) simulations for larger $V$ \cite{Ahmad1998SoluteSolidification,Danilov2006Phase-fieldAlloy,Steinbach2012Phase-fieldDissipation,Kavousi2021QuantitativeSolidification}. Simulating quantitatively far-from-equilibrium conditions, which is relevant for a host of rapid solidification processes, has remained a major challenge.

In this Letter, we develop a PF formulation to quantitatively model dilute alloy solidification under 
far-from-equilibrium conditions with a computationally tractable choice of $W$ on the pattern scale. The model incorporates 
well-known nonequilibrium effects, including solute trapping characterized by $V$-dependent forms of the partition coefficient $k(V)$ and liquidus slope $m(V)$ and interface kinetics. Simulations reproduce the formation of banded microstructures \cite{Boettinger1984TheAlloys,Zimmermann1991CharacterizationAlloys,Gremaud1991BandingTreatment,Carrard1992AboutAlloys,Gill1993RapidlyMap,Gill1995RapidlyMap,Gremaud1990TheTreatment,Kurz1996BandedMicrostructures,McKeown2014InAlloy,McKeown2016Time-ResolvedManufacturing} with a band spacing that is in remarkably good quantitative agreement with observations in thin-films of rapidly solidified Al-Cu alloys \cite{McKeown2016Time-ResolvedManufacturing}. They further reveal that steady-state dendritic array growth is terminated by a novel dendrite tip instability driven by solute trapping that initiates banding. 


\begin{figure}[htbp!]
\includegraphics{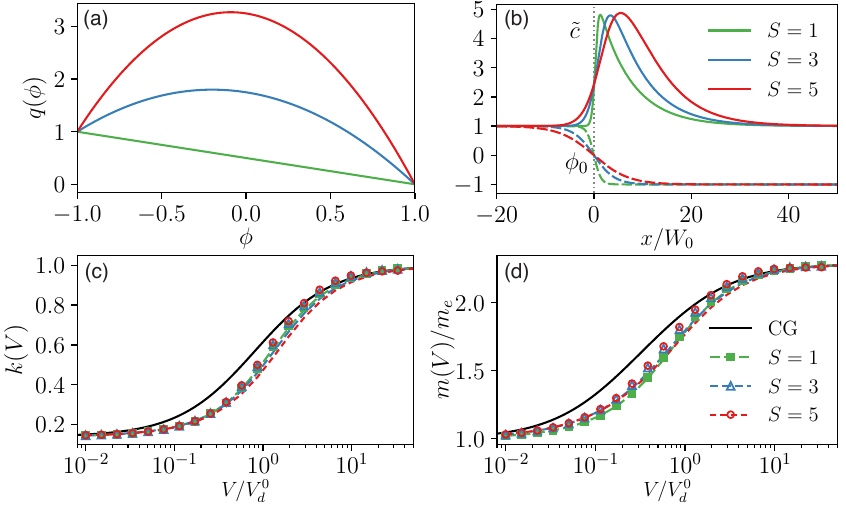}
\caption{\label{fig:figure1} (a) Plots of $q(\phi)$ for $S=1,3,5$ ($A=1,6,12$), with (b) corresponding PF $\phi=\phi_0$ and normalized concentration $\tilde{c}=c/c_\infty$ profiles obtained from the numerical solution of Eq.\ \eqref{c_scale}. (c) $k(V)$ and (d) $m(V)$ functions obtained from the full (symbols) and approximate (dashed lines) solutions (see text). The black solid lines in (c) and (d) represent the CG model with coefficients derived in the large-velocity asymptotic limit \cite{SeeSupplemental}.
}
\end{figure}

PF models have been shown to reproduce solute trapping properties
\cite{Ahmad1998SoluteSolidification,Karma2003Phase-fieldFormation,Danilov2006Phase-fieldAlloy,Steinbach2012Phase-fieldDissipation,Kavousi2021QuantitativeSolidification,Galenko2011SoluteStudy},
quantitatively for a physical choice of interface thickness $W_0 \sim 1$ nm. Computations on a microstructural length scale, however, generally require choosing the interface thickness in the PF model, $W\gg W_0$, thereby producing spurious excess trapping.  
For the low-velocity solidification regime, this problem has been circumvented by the introduction of an antitrapping current that eliminates excess solute trapping to restore local equilibrium at the interface \cite{Karma2001,Echebarria2004}. The form of this current has been modified to also model a moderate departure from equilibrium \cite{Pinomaa2019QuantitativeSolidification,Pinomaa2020PhaseFilms}. However, a quantitative approach remains lacking to describe far-from-equilibrium phenomena such as banding. 
Here, we follow a different approach where excess trapping resulting from the computational constraint $S\equiv W/W_0\gg 1$ is compensated by enhancing the solute diffusivity $D(\phi)\equiv D_lq(\phi)$ in the spatially diffuse interface region. 
We show that, remarkably, simple forms of $q(\phi)$ can be used to reproduce quantitatively the desired velocity-dependent forms of $k(V)$ and $m(V)$ over a several orders of magnitude variation of $V$ from near ($k(V)\rightarrow k_e$ where $k_e$ is the equilibrium value of the partition coefficient) to far from ($k(V)\rightarrow 1$) equilibrium conditions. This approach has the advantage that it can be implemented in a variational formulation of the PF evolution equations that can be readily extended to general binary or multicomponent alloys. 

We present the model for the simplest case of a dilute binary alloy where the evolution equations   
\begin{align}
\tau(\mathbf{n}) \frac{\partial \phi}{\partial t}=& \vec{\nabla} \cdot\left[W(\mathbf{n})^{2} \vec{\nabla} \phi\right] +\phi-\phi^{3} \nonumber\\
&+ \sum_{i=x,y}\left[\partial_{i}\left(|\vec{\nabla} \phi|^{2} W(\mathbf{n}) \frac{\partial W(\mathbf{n})}{\partial\left(\partial_{i} \phi\right)}\right)\right] \nonumber \\
&- \lambda g^{\prime}(\phi) \left[c+ \frac{\left(T-T_{M}\right)}{m_{e}} e^{b(1+g(\phi))}\right], \label{PF_2D_phi} \\
\frac{\partial c}{\partial t}=& \vec{\nabla} \cdot \left\{D_lq(\phi) c \vec{\nabla}[\ln c-b g(\phi)] \right\}, \label{PF_2D_c}
\end{align}
are derived variationally from the free-energy functional introduced in \cite{Karma2003Phase-fieldFormation} and defined here by Eqs.\ (S3)-(S6) in the Supplemental Material \cite{SeeSupplemental}.
Together with the relations $b\equiv \ln k_e/2<0$, $\tau_0=(SW_0)^2/(\Gamma \mu_k^0)$, and $\lambda=a_{1}^0 b m_e SW_0/[\Gamma (k_e-1)]>0$ between PF and materials parameters \cite{Karma2003Phase-fieldFormation}, where $\Gamma=\gamma_0T_M/L$ is the Gibbs-Thomson coefficient, $T_M$ is the melting point, $L$ is the latent heat of melting, $m_e>0$ is the equilibrium value of the liquidus slope, and
$a_1^0=SW_0\int_{-\infty}^{\infty} dx (d\phi_0/dx)^2=2\sqrt{2}/3$, the choices of interface width $W(\mathbf{n})=SW_0a_s(\mathbf{n})$ and time constant 
$\tau(\mathbf{n})=\tau_0a_s(\mathbf{n})^2/a_k(\mathbf{n})$ model general anisotropic forms of the
excess free-energy of the solid-liquid interface $\gamma(\mathbf{n})=\gamma_0a_s(\mathbf{n})$ and interface kinetic coefficient $\mu_k(\mathbf{n})=\mu_k^0a_k(\mathbf{n})$, where $\mathbf{n}$ is the direction normal to the interface. In addition, we use the common choice
$g(\phi)=15( \phi- 2\phi^{3}/3+\phi^{5}/5 )/8$ that satisfies 
$g^{\prime}(\pm 1)=g^{\prime\prime}(\pm 1)=0$ and guarantees that the local minima of the free-energy density remain at $\phi=\pm 1$ for arbitrarily large thermodynamic driving force.

For the one-sided model of alloy solidification, $q(\phi)=(1-\phi)/2$ is the simplest form that describes the physically expected monotonous decrease of diffusivity from liquid to solid across the interface. This form, however, produces spurious excess trapping at lower $V$ when $S\gg 1$, since the diffusive speed in the PF model $V_d\sim D_l/W\sim V_d^0/S$, where $V_d^0\equiv D_l/W_0$. Hence, to eliminate excess trapping, we use the quadratic form $q(\phi)=A(1-\phi)/2-(A-1)(1-\phi)^2/4$ that enhances $D(\phi)$ in the interface region for $A>1$ (Fig.\ \ref{fig:figure1}(a)). 


We show next how this form of $q(\phi)$ can be used to reproduce $S$-independent solute trapping properties.
For this, we look for steady-state PF and concentration profiles corresponding to a planar isothermal interface moving at constant velocity $V$. Those profiles are determined by rewriting Eqs.\ \eqref{PF_2D_phi}-\eqref{PF_2D_c} in a frame moving with the interface at velocity $V$ in the $x$ direction
\begin{align}
-\tau_0 V \frac{d \phi}{d x} &=(SW_0)^{2} \frac{d^2 \phi}{dx^2} + \phi - \phi^3 \nonumber \\
-&\lambda g^{\prime}(\phi)\left[c+\frac{(T-T_M)}{m_e} e^{b(1+g(\phi))}\right], \label{eq1} \\
- V \frac{dc}{dx}&= \frac{d}{dx} \left(D_l q(\phi) c \frac{d}{dx}[\ln c-b g(\phi)] \right), \label{eq2}
\end{align} 
where we have considered for simplicity the isotropic case
$a_s(\mathbf{n})=a_k(\mathbf{n})=1$. Eq.\ \eqref{eq2} can be simplified further by integrating both sides once with respect to $x$ and using the boundary condition $c(\pm \infty)=c_\infty$ imposed by mass conservation, yielding
\begin{equation}
\frac{d c}{d x}= (c_\infty-c) \frac{V}{D_l q(\phi)}+b\, c\, \frac{d g(\phi)}{d x}. \label{c_scale}
\end{equation}  
In addition, a
self-consistent expression for the velocity-dependent temperature is obtained by multiplying both sides of Eq.\ \eqref{eq1} by $d\phi/dx$ and integrating over $x$ from $-\infty$ to $+\infty$, yielding
\begin{equation}
T(V)=T_{M}- \frac{b\, m_e}{1-k_e} \int_{-\infty}^{\infty} dx\,g^{\prime}(\phi) \,c\, \frac{d\phi}{dx}+\frac{a_1 b \,m_e}{\lambda(1-k_e)} \frac{\tau_0 V}{SW_0}, \label{T_PF}
\end{equation}
where we have defined $a_1=SW_0\int_{-\infty}^{\infty} dx (d\phi/dx)^2$. The solution of Eqs.\ \eqref{eq1} and \eqref{c_scale} with $T$ given by Eq.\ \eqref{T_PF} and the boundary condition $c(\pm \infty) =c_{\infty}$ uniquely determine the steady-state profile $\phi(x)$ and $c(x)$. The ``full solution'' to this system of equations is straightforward to obtain numerically by a procedure that will be described in more detail elsewhere. An ``approximate solution'' very close to the full solution can also be obtained by assuming that the PF profile for a moving interface remains close to its stationary profile $\phi_0(x)=-\tanh{ [ {x}/{(\sqrt{2} W}) ]}$. In this approximation, the concentration profile is solely determined by Eq.\ \eqref{c_scale}, which is readily solved by numerical integration to obtain the concentration profiles shown in Fig. \ref{fig:figure1}(b). $T(V)$ is then determined by Eq.\ \eqref{T_PF} with those $c$ profiles and $\phi(x)=\phi_0(x)$. 
Finally, the corresponding functions $k(V)$ and $m(V)$ are obtained from the sharp-interface relations $k(V)=c_s/c_l$ and
\begin{equation}
T(V)=T_{M}-m(V) c_{l} - {V}/{\mu_k}, \label{Gibbs_Thomson}
\end{equation}
where $c_s$ and $c_l$ are the concentrations on the solid and liquid sides of the interface, respectively, which correspond here to $c_\infty$ and the peak value of $c(x)$. Matching the second and third terms on the
right-hand side of Eqs.\ \eqref{T_PF} and \eqref{Gibbs_Thomson}, yields at once
\begin{equation}
\frac{m(V)}{m_e}=\frac{b}{(1-k_e)c_l} \int_{-\infty}^{\infty} dx g^{\prime}(\phi) c \frac{d\phi}{dx}, \label{m_v_m_e}
\end{equation}
and $\mu_k=\mu_k^0 a_1^0/a_1$, respectively. 
To obtain values of $A$ that yield $S$-independent trapping properties, we first compute reference $k(V)$ and $m(V)$ curves corresponding to $S=1$ and $A=1$. For a given $S>1$, we then compute $k(V)$ and $m(V)$ curves for different $A$ values and find the value of $A$ that minimizes the departure from the reference curves over some large velocity range of interest. This procedure is implemented with the approximate ($\phi=\phi_0$) solution and yields
$A=6$ and 12 for $S=3$ and 5, respectively, for parameters of Al-Cu alloys \cite{SeeSupplemental}.      
Plots of $k(V)$ and $m(V)$ obtained from the approximate ($\phi=\phi_0$) and full solutions of the steady-state concentration and PF profiles are shown for the different $S$ and corresponding $A$ values in Figs.\ \ref{fig:figure1}(c)-(d), respectively. The approximate solution only depends on $k_e$ and yields $\mu_k=\mu_k^0$ while the full solution depends on the other alloy parameters, and the deviation of $\phi$ from $\phi_0$ at larger $V/V_d^0$ causes a small quantitative difference between the two solutions. Remarkably, even though a single parameter $A$ is optimized for each $S$, $k(V)$ and $m(V)$ are seen to be nearly independent of $S$ over a several orders magnitude variation of $V$. Even though the concentration profiles depend on $S$ (Fig. \ref{fig:figure1}(b)), they have almost identical peak values, which determine $k(V)$, and the different profiles also yield nearly identical values of $m(V)$ via Eq.\ \eqref{m_v_m_e}. 

The $k(V)$ and $m(V)$ functions are compared to the predictions of the continuous growth (CG) model as in \cite{Ahmad1998SoluteSolidification} by extracting the diffusive speed $V_d$ from the asymptotic analytical solution of Eq.\ \eqref{c_scale} for $V\gg V_d^0$ and $S=1$, assuming $\phi=\phi_0$, and concomitantly the solute drag coefficient $\alpha$ from Eq.\ \eqref{m_v_m_e}. This calculation yields $V_d\approx 0.356 V_d^0 \ln(1/k_e)/(1-k_e)$ and $\alpha\approx 0.645$ \cite{SeeSupplemental}. Since the PF model resolves the spatially diffuse interface region, while the CG model is a sharp-interface description, quantitative agreement between the two models over the entire range of $V$ for different $k_e$ is not generally expected, even though agreement becomes almost perfect for larger $k_e$ \cite{SeeSupplemental}. More than the PF and CG models comparison, what is important here is that the PF $k(V)$ and $m(V)$ curves for a realistic width $(S=1)$ can be reproduced for a much larger width ($S\gg 1$) to make simulations on a microstructural scale feasible. Parameters of the PF model (e.g., $W_0$ and the functions $q(\phi)$ and $g(\phi)$) can, in addition, be further adjusted to better fit desired $k(V)$ and $m(V)$ curves. 

\begin{figure}[htbp!]
\includegraphics{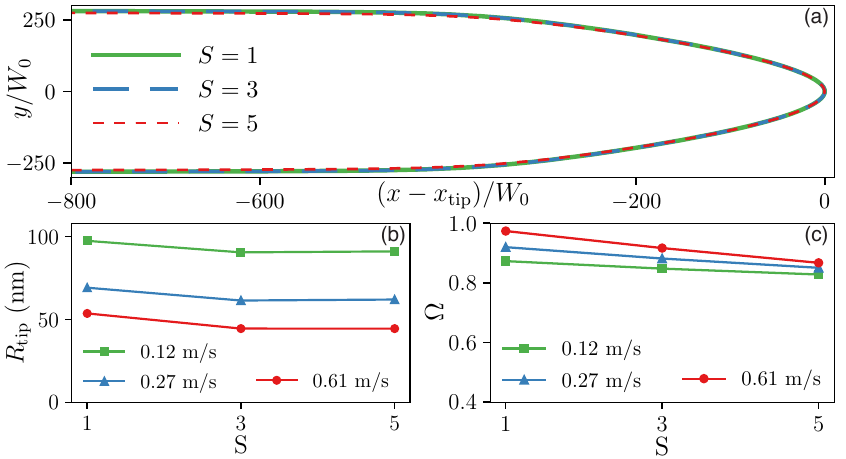}
\caption{\label{fig:figure2} 
(a) Comparison of steady-state interface shapes for different $S$ from 2D PF simulations of dendritic array growth for Al-3wt.\% Cu, $V_p=0.12$ m/s, and $G=5\times 10^6$ K/m. (b) Tip radius $R_{\mathrm{tip}}$ and (c) dimensionless tip supersaturation $\Omega$ versus $S$ for different $V_p$.
}
\end{figure}

To validate our approach for such simulations, we model the two-dimensional (2D) directional solidification of dilute Al-Cu alloys. We consider first the standard frozen temperature approximation (FTA) that neglects latent-heat rejection, which corresponds to replacing ${\left(T-T_{M}\right)}/{m_{e}}$ in Eq.\ \eqref{PF_2D_phi} by $ G(x-x_0-V_p t)/m_e-c_{\infty}$, where $V_p$ is the pulling speed of the sample, $G$ is the externally imposed temperature gradient, and $x_0$ coincides with the equilibrium liquidus temperature $T_L-m_ec_{\infty}$. In addition, we consider anisotropic forms of the excess interface free-energy $a_s(\theta)=1+\epsilon_s \cos(4\theta)$, and kinetic coefficient $a_k(\theta)=1+\epsilon_k \cos(4\theta)$, with fourfold symmetry, where $\theta$ is the angle between $\mathbf{n}$ and the $x$ axis.

To investigate the convergence of the method, we first focus on the velocity range below the onset of banding where stable dendritic array structures are formed. This allows us to compare, for different $S$, steady-state interface shapes corresponding to a single dendrite obtained using periodic boundary conditions in $y$, with the width of the simulation domain along $y$ equal to the primary dendrite array spacing $\Lambda$ (0.65 $\mu$m), chosen within the stable range of $\Lambda$. This comparison in Fig.\ \ref{fig:figure2}(a) shows that different $S$ yield nearly identical shapes, and the computation time is reduced by three orders of magnitude for $S=5$ compared to $S=1$ \cite{SeeSupplemental}. Results in Fig.\ \ref{fig:figure2}(b) characterize steady-state shapes by the tip radius $R_{\mathrm{tip}}$ and dimensionless tip supersaturation $\Omega=(c_l-c_\infty)/(c_l-c_s)$. The latter is relatively well described by the Ivantsov relation between $\Omega$ and P\'eclet number $R_{\mathrm{tip}}V_p/(2D_l)$ \cite{SeeSupplemental}. Quantitative differences between different $S$ for larger $V_p$ are likely due to other effects, such as surface diffusion and interface stretching known to affect pattern selection \cite{Karma2001,Echebarria2004}. While those effects can be eliminated in the framework of the thin-interface limit for quasi-equilibrium growth conditions, eliminating them for the entire $V$ range of Figs.\ \ref{fig:figure1}(c)-(d) is considerably more challenging. Fig.\ \ref{fig:figure2} shows that, even with such effects present, the method converges already reasonably well. 

\begin{figure}[htbp!]
\includegraphics[scale=1]{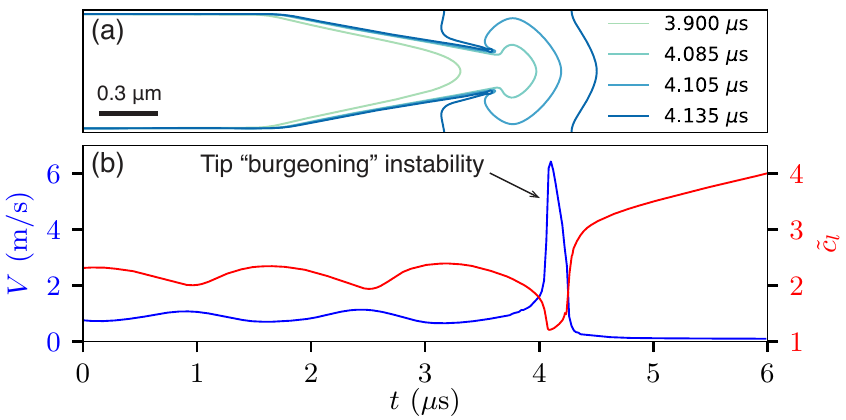}
\caption{\label{fig:figure3} (a) Evolution of the solid-liquid interface illustrating the burgeoning tip instability at $V_p=V_c=0.88$ m/s for Al-3wt.\% Cu and $G=5\times 10^6$ K/m (see movie in \cite{SeeSupplemental}). (b) Corresponding tip velocity $V$ and scaled interface concentration $\tilde{c}_l=c_l/c_\infty$ on the liquid side. 
}
\end{figure}

Next, we exploit the model to address basic open questions of interface dynamics in this regime using $S=5$ for efficiency. The first is how steady-state dendrite array growth illustrated in Fig.\ \ref{fig:figure2}(a) loses stability to trigger banding as $V_p$ approaches the absolute stability limit $V_a$ defined implicitly by the relation
\cite{Mullins1964StabilityAlloy,Trivedi1986MorphologicalConditions,Ludwig1996DirectStability,Boettinger1999SimulationVelocity}
\begin{equation}
V_a=\frac{D_l m(V_a)c_{\infty}\left[1-k(V_a)\right]}{k(V_a)^{2} \Gamma}, \label{Va}
\end{equation}
where $k(V)$ and $m(V)$ are computed as before from the full solution of the 1D PF Eqs.\ \eqref{eq1} and \eqref{c_scale} with $T(V)$ given by Eq.\ \eqref{T_PF}, but with the substitutions $SW_0\rightarrow SW_0(1+\epsilon_s)$ and $\tau_0\rightarrow \tau_0(1+\epsilon_k)$ in Eq.\ \eqref{eq1} to account for anisotropy. To address this question, we performed a simulation in the same geometry Fig.\ \ref{fig:figure2}(a) but with $V_p$ slowly increasing linearly in time on a timescale much longer than the characteristic time for the interface to relax to a steady-state shape, thereby allowing us to probe pattern stability over a large range of $V_p$. We find that above a critical velocity $V_c\approx 0.88$ m/s steady-state growth becomes unstable as illustrated by the time sequence in Fig.\ \ref{fig:figure3}(a). This instability is highly localized at the dendrite tip and triggers a rapid ``burgeoning-like'' growth of the interface. Fig.\ \ref{fig:figure3}(b) shows that this abrupt acceleration of the interface is accompanied by a rapid drop in $c_l$ associated with almost complete solute trapping, followed by a rapid deceleration of the interface and increase of $c_l$ as the interface transits to a planar morphology and the diffusion boundary layer rebuilds itself. 

The onset velocity $V_c$ of instability depends on the anisotropy parameters $\epsilon_s$ and $\epsilon_k$ that are known to control dendrite tip selection \cite{Langer1986SolvabilityAnisotropy,Brener1990EffectsDendrites,Brener1991PatternGrowth,Bragard2002} and do not enter in the linear stability analysis used to predict $V_a$. However, for the parameters of our simulations, $V_c$ turns out to be close to the value $V_a\approx 0.86$ m/s predicted by Eq.\ \eqref{Va}. Moreover, the simulation of Fig.\ \ref{fig:figure3} was purposely carried out without thermal noise to study the basic instability of steady-state shapes. Additional simulations with noise-induced sidebranching reveal that the burgeoning instability can also emerge from the tips of secondary branches, especially for larger $\Lambda$ that accommodates larger amplitude sidebranches.

\begin{figure}[htbp!]
\includegraphics[scale=1]{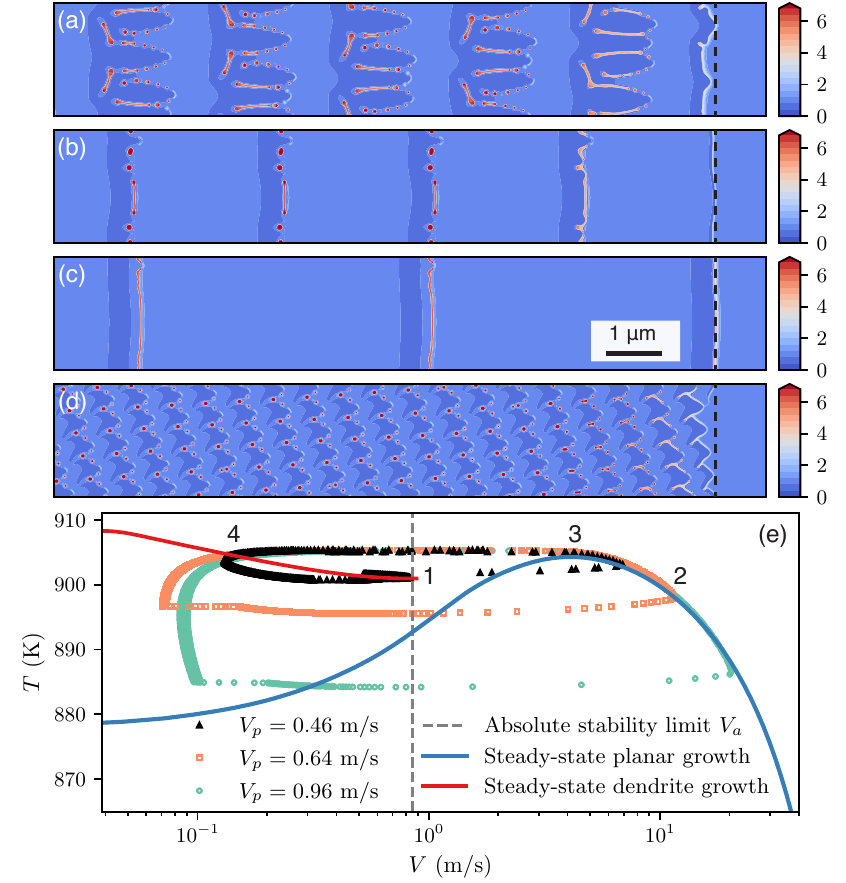}
\caption{\label{fig:figure4} 
Microsegregation patterns ($\tilde{c}=c/c_\infty$ color maps) simulated for Al-3wt.\% Cu with $G=5\times 10^6$ K/m and the FTA for $V_p=$ (a) 0.46, (b) 0.64, and (c) 0.96 m/s, and with latent heat for (d) $V_p=0.96 $ m/s; black dashed lines mark solidification front positions. (e) Steady-state curves and banding cycles in the $T-V$ plane. Movies corresponding to (a) and (d) are shown in the Supplemental Material \cite{SeeSupplemental}. 
}
\end{figure}

Next, we investigate banding
by using the FTA and the method developed in \cite{Song2018Thermal-fieldSolidification} to include latent-heat rejection at the interface. Figs.\ \ref{fig:figure4}(a)-(c) show the microstructures obtained with the FTA at three increasing values of $V_p$ and Fig.\ \ref{fig:figure4}(d) shows the pattern obtained at the largest $V_p$ with latent heat for comparison. The oscillation cycles corresponding to the FTA simulations of Figs.\ \ref{fig:figure4}(a)-(c) are shown in the $T-V$ plane of Fig.\ \ref{fig:figure4}(e), where points along each cycle represent the temperature and velocity of the most advanced point along the solidification front at subsequent instants of time. 

Superimposed on the 
$T-V$ plane are the steady-state curves corresponding to stable dendritic array growth (red curve) for $V<V_c$ and planar front growth (blue curve) computed using Eq.\ \eqref{Gibbs_Thomson}. The conceptual model of banding derived in the FTA framework assumes that the interface makes instantaneous transitions (1-2) and (3-4) between those steady-state curves, and follows those curves during the dendritic array (4-1) and planar front (2-3) growth portions of the complete 1-2-3-4-1 banding cycle, where 1 corresponds to $V_a$ and 3 corresponds to the maximum of the $T(V)$ curve for steady-state planar front growth. 
The simulated banding cycle of Fig.\ \ref{fig:figure4}(a) follows reasonably well this conceptual cycle when   microsegregated and microsegregation-free bands corresponding to dendritic array and planar front growth, respectively, are of comparable width, while the cycles of Figs.\ \ref{fig:figure4}(b)-(c) make larger loops in the $T-V$ plane when planar front growth occupies a larger fraction of the whole banding cycle that is no longer constrained to follow the 4-1 segment corresponding to  steady-state dendritic array growth. 

The comparison of Fig.\ \ref{fig:figure4} (c) and (d) shows that latent-heat rejection dramatically reduces the band spacing from about 2 $\mu$m to 500 nm. Latent heat was previously found to reduce the period of oscillations of the planar interface  \cite{Karma1992DynamicsSolidification,Karma1993InterfaceSolidification}, but those 1D cycles could not predict banded microstructure formation. 
Fig. \ref{fig:figure4}(d) reveals that bands grow at a small angle with respect to the thermal axis due to the fact that the lateral spreading velocity of the interface that produces microsegregation-free bands is slowed down by latent-heat rejection. The banding cycle is shrunk and no longer easily represented by the path of a uniquely defined solidification front in the $T-V$ plane as for FTA. 

\begin{figure}[htbp!]
\includegraphics[scale=1]{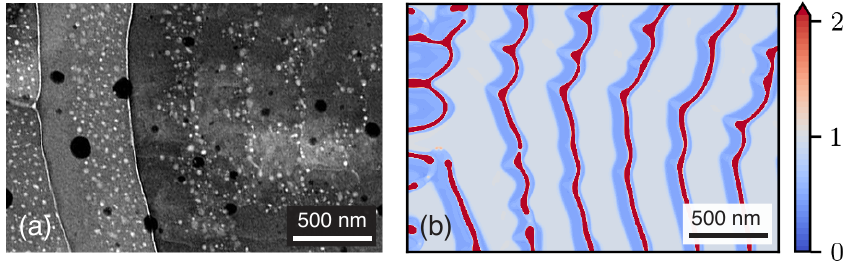}
\caption{\label{fig:figure5} 
Comparison of banded microstructures for Al-9 wt.\% Cu from (a) the late stage of a thin-film resolidification experiment (reproduced with permission from \cite{McKeown2016Time-ResolvedManufacturing}), and (b) a 2D PF simulation with latent heat and $G=5\times 10^6$ K/m ($\tilde{c}=c/c_\infty$ color map).
}
\end{figure}

Finally, we show in Fig.\ \ref{fig:figure5} a quantitative comparison of banded microstructures simulated with latent heat and produced in a resolidification experiment where a short laser pulse is used to create an elliptical melt pool in a thin film of an Al-9 wt.\% (Al-4 at.\%) Cu alloy \cite{McKeown2016Time-ResolvedManufacturing}. Based on dynamic transmission electron microscopy (DTEM) measurements of interface velocity, $V_p$ was increased in the simulation linearly from 0.3 to 1.8 m/s over a time period 30 $\mathrm{\mu s}$. As shown in Fig.\ \ref{fig:figure5}, the band spacing in simulation ($\sim 400$ nm) agrees remarkably well with the experiment. Simulations for other alloys (e.g., Al-Fe) also yield a good agreement with previous experimental observations of banded microstructures in laser remelting experiments \cite{Boettinger1984TheAlloys,Zimmermann1991CharacterizationAlloys,Gremaud1991BandingTreatment,Carrard1992AboutAlloys,Gill1993RapidlyMap,Gill1995RapidlyMap,Gremaud1990TheTreatment}. They will be presented elsewhere in a longer exposition of methods and results.


We thank Wilfried Kurz for valuable discussions. This work was supported by the U.S. Department of Energy (DOE), Office of Science, Basic Energy Sciences (BES) under Award No.\ DE-SC0020870.

\nocite{Lee2004Diffusion-coefficientAlloys,Gunduz1985TheSystems,Mendelev2010Molecular-dynamicsMetals,Ji2022IsotropicSolidification,Aziz1982ModelSolidification,Aziz1988ContinuousSolidification,Aziz1994OnSolidification,Ivantsov1947TemperatureMelt}

\bibliography{references}

\begin{thebibliography}{57}%
\makeatletter
\providecommand \@ifxundefined [1]{%
 \@ifx{#1\undefined}
}%
\providecommand \@ifnum [1]{%
 \ifnum #1\expandafter \@firstoftwo
 \else \expandafter \@secondoftwo
 \fi
}%
\providecommand \@ifx [1]{%
 \ifx #1\expandafter \@firstoftwo
 \else \expandafter \@secondoftwo
 \fi
}%
\providecommand \natexlab [1]{#1}%
\providecommand \enquote  [1]{``#1''}%
\providecommand \bibnamefont  [1]{#1}%
\providecommand \bibfnamefont [1]{#1}%
\providecommand \citenamefont [1]{#1}%
\providecommand \href@noop [0]{\@secondoftwo}%
\providecommand \href [0]{\begingroup \@sanitize@url \@href}%
\providecommand \@href[1]{\@@startlink{#1}\@@href}%
\providecommand \@@href[1]{\endgroup#1\@@endlink}%
\providecommand \@sanitize@url [0]{\catcode `\\12\catcode `\$12\catcode
  `\&12\catcode `\#12\catcode `\^12\catcode `\_12\catcode `\%12\relax}%
\providecommand \@@startlink[1]{}%
\providecommand \@@endlink[0]{}%
\providecommand \url  [0]{\begingroup\@sanitize@url \@url }%
\providecommand \@url [1]{\endgroup\@href {#1}{\urlprefix }}%
\providecommand \urlprefix  [0]{URL }%
\providecommand \Eprint [0]{\href }%
\providecommand \doibase [0]{https://doi.org/}%
\providecommand \selectlanguage [0]{\@gobble}%
\providecommand \bibinfo  [0]{\@secondoftwo}%
\providecommand \bibfield  [0]{\@secondoftwo}%
\providecommand \translation [1]{[#1]}%
\providecommand \BibitemOpen [0]{}%
\providecommand \bibitemStop [0]{}%
\providecommand \bibitemNoStop [0]{.\EOS\space}%
\providecommand \EOS [0]{\spacefactor3000\relax}%
\providecommand \BibitemShut  [1]{\csname bibitem#1\endcsname}%
\let\auto@bib@innerbib\@empty
\bibitem [{\citenamefont {Boettinger}\ \emph {et~al.}(2002)\citenamefont
  {Boettinger}, \citenamefont {Warren}, \citenamefont {Beckermann},\ and\
  \citenamefont {Karma}}]{Boettinger2002Phase-fieldSolidification}%
  \BibitemOpen
  \bibfield  {author} {\bibinfo {author} {\bibfnamefont {W.~J.}\ \bibnamefont
  {Boettinger}}, \bibinfo {author} {\bibfnamefont {J.~A.}\ \bibnamefont
  {Warren}}, \bibinfo {author} {\bibfnamefont {C.}~\bibnamefont {Beckermann}},\
  and\ \bibinfo {author} {\bibfnamefont {A.}~\bibnamefont {Karma}},\ }\href
  {https://doi.org/10.1146/annurev.matsci.32.101901.155803} {\bibinfo {title}
  {{Phase-field simulation of solidification}}} (\bibinfo {year}
  {2002})\BibitemShut {NoStop}%
\bibitem [{\citenamefont {Steinbach}(2009)}]{Steinbach2009Phase-fieldScience}%
  \BibitemOpen
  \bibfield  {author} {\bibinfo {author} {\bibfnamefont {I.}~\bibnamefont
  {Steinbach}},\ }\href {https://doi.org/10.1088/0965-0393/17/7/073001}
  {\bibfield  {journal} {\bibinfo  {journal} {Modelling and Simulation in
  Materials Science and Engineering}\ }\textbf {\bibinfo {volume} {17}},\
  \bibinfo {pages} {073001} (\bibinfo {year} {2009})}\BibitemShut {NoStop}%
\bibitem [{\citenamefont {Karma}\ and\ \citenamefont
  {Tourret}(2016)}]{Karma2016}%
  \BibitemOpen
  \bibfield  {author} {\bibinfo {author} {\bibfnamefont {A.}~\bibnamefont
  {Karma}}\ and\ \bibinfo {author} {\bibfnamefont {D.}~\bibnamefont
  {Tourret}},\ }\href {https://doi.org/10.1016/j.cossms.2015.09.001} {\bibinfo
  {title} {{Atomistic to continuum modeling of solidification
  microstructures}}} (\bibinfo {year} {2016})\BibitemShut {NoStop}%
\bibitem [{\citenamefont {Kurz}\ \emph {et~al.}(2020)\citenamefont {Kurz},
  \citenamefont {Rappaz},\ and\ \citenamefont
  {Trivedi}}]{Kurz2020Progress2018}%
  \BibitemOpen
  \bibfield  {author} {\bibinfo {author} {\bibfnamefont {W.}~\bibnamefont
  {Kurz}}, \bibinfo {author} {\bibfnamefont {M.}~\bibnamefont {Rappaz}},\ and\
  \bibinfo {author} {\bibfnamefont {R.}~\bibnamefont {Trivedi}},\ }\href
  {https://doi.org/10.1080/09506608.2020.1757894} {\bibfield  {journal}
  {\bibinfo  {journal} {International Materials Reviews}\ }\textbf {\bibinfo
  {volume} {66}},\ \bibinfo {pages} {30} (\bibinfo {year} {2020})}\BibitemShut
  {NoStop}%
\bibitem [{\citenamefont {Tourret}\ \emph {et~al.}(2022)\citenamefont
  {Tourret}, \citenamefont {Liu},\ and\ \citenamefont
  {LLorca}}]{Tourret2022Phase-fieldChallenges}%
  \BibitemOpen
  \bibfield  {author} {\bibinfo {author} {\bibfnamefont {D.}~\bibnamefont
  {Tourret}}, \bibinfo {author} {\bibfnamefont {H.}~\bibnamefont {Liu}},\ and\
  \bibinfo {author} {\bibfnamefont {J.}~\bibnamefont {LLorca}},\ }\href
  {https://doi.org/10.1016/J.PMATSCI.2021.100810} {\bibfield  {journal}
  {\bibinfo  {journal} {Progress in Materials Science}\ }\textbf {\bibinfo
  {volume} {123}},\ \bibinfo {pages} {100810} (\bibinfo {year}
  {2022})}\BibitemShut {NoStop}%
\bibitem [{\citenamefont {Karma}\ and\ \citenamefont
  {Rappel}(1998)}]{Karma1998}%
  \BibitemOpen
  \bibfield  {author} {\bibinfo {author} {\bibfnamefont {A.}~\bibnamefont
  {Karma}}\ and\ \bibinfo {author} {\bibfnamefont {W.~J.}\ \bibnamefont
  {Rappel}},\ }\href {https://doi.org/10.1103/PhysRevE.57.4323} {\bibfield
  {journal} {\bibinfo  {journal} {Physical Review E - Statistical Physics,
  Plasmas, Fluids, and Related Interdisciplinary Topics}\ ,\ \bibinfo {pages}
  {4323}} (\bibinfo {year} {1998})}\BibitemShut {NoStop}%
\bibitem [{\citenamefont {Karma}(2001)}]{Karma2001}%
  \BibitemOpen
  \bibfield  {author} {\bibinfo {author} {\bibfnamefont {A.}~\bibnamefont
  {Karma}},\ }\href {https://doi.org/10.1103/PhysRevLett.87.115701} {\bibfield
  {journal} {\bibinfo  {journal} {Physical Review Letters}\ ,\ \bibinfo {pages}
  {115701}} (\bibinfo {year} {2001})}\BibitemShut {NoStop}%
\bibitem [{\citenamefont {Echebarria}\ \emph {et~al.}(2004)\citenamefont
  {Echebarria}, \citenamefont {Folch}, \citenamefont {Karma},\ and\
  \citenamefont {Plapp}}]{Echebarria2004}%
  \BibitemOpen
  \bibfield  {author} {\bibinfo {author} {\bibfnamefont {B.}~\bibnamefont
  {Echebarria}}, \bibinfo {author} {\bibfnamefont {R.}~\bibnamefont {Folch}},
  \bibinfo {author} {\bibfnamefont {A.}~\bibnamefont {Karma}},\ and\ \bibinfo
  {author} {\bibfnamefont {M.}~\bibnamefont {Plapp}},\ }\href
  {https://doi.org/10.1103/PhysRevE.70.061604} {\bibfield  {journal} {\bibinfo
  {journal} {Physical Review E - Statistical Physics, Plasmas, Fluids, and
  Related Interdisciplinary Topics}\ ,\ \bibinfo {pages} {061604}} (\bibinfo
  {year} {2004})}\BibitemShut {NoStop}%
\bibitem [{\citenamefont {Folch}\ and\ \citenamefont
  {Plapp}(2005)}]{Folch2005QuantitativeGrowth}%
  \BibitemOpen
  \bibfield  {author} {\bibinfo {author} {\bibfnamefont {R.}~\bibnamefont
  {Folch}}\ and\ \bibinfo {author} {\bibfnamefont {M.}~\bibnamefont {Plapp}},\
  }\href {https://doi.org/10.1103/PHYSREVE.72.011602/FIGURES/18/MEDIUM}
  {\bibfield  {journal} {\bibinfo  {journal} {Physical Review E - Statistical,
  Nonlinear, and Soft Matter Physics}\ }\textbf {\bibinfo {volume} {72}},\
  \bibinfo {pages} {011602} (\bibinfo {year} {2005})}\BibitemShut {NoStop}%
\bibitem [{\citenamefont {Plapp}(2011)}]{Plapp2011UnifiedFunctional}%
  \BibitemOpen
  \bibfield  {author} {\bibinfo {author} {\bibfnamefont {M.}~\bibnamefont
  {Plapp}},\ }\href
  {https://doi.org/10.1103/PHYSREVE.84.031601/FIGURES/1/MEDIUM} {\bibfield
  {journal} {\bibinfo  {journal} {Physical Review E - Statistical, Nonlinear,
  and Soft Matter Physics}\ }\textbf {\bibinfo {volume} {84}},\ \bibinfo
  {pages} {031601} (\bibinfo {year} {2011})}\BibitemShut {NoStop}%
\bibitem [{\citenamefont {Boussinot}\ and\ \citenamefont
  {Brener}(2014)}]{Boussinot2014AchievingContrast}%
  \BibitemOpen
  \bibfield  {author} {\bibinfo {author} {\bibfnamefont {G.}~\bibnamefont
  {Boussinot}}\ and\ \bibinfo {author} {\bibfnamefont {E.~A.}\ \bibnamefont
  {Brener}},\ }\href
  {https://doi.org/10.1103/PHYSREVE.89.060402/FIGURES/1/MEDIUM} {\bibfield
  {journal} {\bibinfo  {journal} {Physical Review E - Statistical, Nonlinear,
  and Soft Matter Physics}\ }\textbf {\bibinfo {volume} {89}},\ \bibinfo
  {pages} {060402} (\bibinfo {year} {2014})}\BibitemShut {NoStop}%
\bibitem [{\citenamefont {Haxhimali}\ \emph {et~al.}(2006)\citenamefont
  {Haxhimali}, \citenamefont {Karma}, \citenamefont {Gonzales},\ and\
  \citenamefont {Rappaz}}]{Haxhimali2006OrientationEvolution}%
  \BibitemOpen
  \bibfield  {author} {\bibinfo {author} {\bibfnamefont {T.}~\bibnamefont
  {Haxhimali}}, \bibinfo {author} {\bibfnamefont {A.}~\bibnamefont {Karma}},
  \bibinfo {author} {\bibfnamefont {F.}~\bibnamefont {Gonzales}},\ and\
  \bibinfo {author} {\bibfnamefont {M.}~\bibnamefont {Rappaz}},\ }\bibfield
  {journal} {\bibinfo  {journal} {Nature Materials}\ }\href
  {https://doi.org/10.1038/nmat1693} {10.1038/nmat1693} (\bibinfo {year}
  {2006})\BibitemShut {NoStop}%
\bibitem [{\citenamefont {Dantzig}\ \emph {et~al.}(2013)\citenamefont
  {Dantzig}, \citenamefont {Di~Napoli}, \citenamefont {Friedli},\ and\
  \citenamefont {Rappaz}}]{Dantzig2013DendriticComputations}%
  \BibitemOpen
  \bibfield  {author} {\bibinfo {author} {\bibfnamefont {J.~A.}\ \bibnamefont
  {Dantzig}}, \bibinfo {author} {\bibfnamefont {P.}~\bibnamefont {Di~Napoli}},
  \bibinfo {author} {\bibfnamefont {J.}~\bibnamefont {Friedli}},\ and\ \bibinfo
  {author} {\bibfnamefont {M.}~\bibnamefont {Rappaz}},\ }\href
  {https://doi.org/10.1007/S11661-013-1911-8/FIGURES/8} {\bibfield  {journal}
  {\bibinfo  {journal} {Metallurgical and Materials Transactions A: Physical
  Metallurgy and Materials Science}\ }\textbf {\bibinfo {volume} {44}},\
  \bibinfo {pages} {5532} (\bibinfo {year} {2013})}\BibitemShut {NoStop}%
\bibitem [{\citenamefont {Bergeon}\ \emph {et~al.}(2013)\citenamefont
  {Bergeon}, \citenamefont {Tourret}, \citenamefont {Chen}, \citenamefont
  {Debierre}, \citenamefont {Gu{\'{e}}rin}, \citenamefont {Ramirez},
  \citenamefont {Billia}, \citenamefont {Karma},\ and\ \citenamefont
  {Trivedi}}]{Bergeon2013SpatiotemporalSolidification}%
  \BibitemOpen
  \bibfield  {author} {\bibinfo {author} {\bibfnamefont {N.}~\bibnamefont
  {Bergeon}}, \bibinfo {author} {\bibfnamefont {D.}~\bibnamefont {Tourret}},
  \bibinfo {author} {\bibfnamefont {L.}~\bibnamefont {Chen}}, \bibinfo {author}
  {\bibfnamefont {J.~M.}\ \bibnamefont {Debierre}}, \bibinfo {author}
  {\bibfnamefont {R.}~\bibnamefont {Gu{\'{e}}rin}}, \bibinfo {author}
  {\bibfnamefont {A.}~\bibnamefont {Ramirez}}, \bibinfo {author} {\bibfnamefont
  {B.}~\bibnamefont {Billia}}, \bibinfo {author} {\bibfnamefont
  {A.}~\bibnamefont {Karma}},\ and\ \bibinfo {author} {\bibfnamefont
  {R.}~\bibnamefont {Trivedi}},\ }\href
  {https://doi.org/10.1103/PhysRevLett.110.226102} {\bibfield  {journal}
  {\bibinfo  {journal} {Physical Review Letters}\ ,\ \bibinfo {pages} {226102}}
  (\bibinfo {year} {2013})}\BibitemShut {NoStop}%
\bibitem [{\citenamefont {Clarke}\ \emph {et~al.}(2017)\citenamefont {Clarke},
  \citenamefont {Tourret}, \citenamefont {Song}, \citenamefont {Imhoff},
  \citenamefont {Gibbs}, \citenamefont {Gibbs}, \citenamefont {Fezzaa},\ and\
  \citenamefont {Karma}}]{Clarke2017}%
  \BibitemOpen
  \bibfield  {author} {\bibinfo {author} {\bibfnamefont {A.~J.}\ \bibnamefont
  {Clarke}}, \bibinfo {author} {\bibfnamefont {D.}~\bibnamefont {Tourret}},
  \bibinfo {author} {\bibfnamefont {Y.}~\bibnamefont {Song}}, \bibinfo {author}
  {\bibfnamefont {S.~D.}\ \bibnamefont {Imhoff}}, \bibinfo {author}
  {\bibfnamefont {P.~J.}\ \bibnamefont {Gibbs}}, \bibinfo {author}
  {\bibfnamefont {J.~W.}\ \bibnamefont {Gibbs}}, \bibinfo {author}
  {\bibfnamefont {K.}~\bibnamefont {Fezzaa}},\ and\ \bibinfo {author}
  {\bibfnamefont {A.}~\bibnamefont {Karma}},\ }\bibfield  {journal} {\bibinfo
  {journal} {Acta Materialia}\ }\href
  {https://doi.org/10.1016/j.actamat.2017.02.047}
  {10.1016/j.actamat.2017.02.047} (\bibinfo {year} {2017})\BibitemShut
  {NoStop}%
\bibitem [{\citenamefont {Song}\ \emph {et~al.}(2018)\citenamefont {Song},
  \citenamefont {Tourret}, \citenamefont {Mota}, \citenamefont {Pereda},
  \citenamefont {Billia}, \citenamefont {Bergeon}, \citenamefont {Trivedi},\
  and\ \citenamefont {Karma}}]{Song2018Thermal-fieldSolidification}%
  \BibitemOpen
  \bibfield  {author} {\bibinfo {author} {\bibfnamefont {Y.}~\bibnamefont
  {Song}}, \bibinfo {author} {\bibfnamefont {D.}~\bibnamefont {Tourret}},
  \bibinfo {author} {\bibfnamefont {F.~L.}\ \bibnamefont {Mota}}, \bibinfo
  {author} {\bibfnamefont {J.}~\bibnamefont {Pereda}}, \bibinfo {author}
  {\bibfnamefont {B.}~\bibnamefont {Billia}}, \bibinfo {author} {\bibfnamefont
  {N.}~\bibnamefont {Bergeon}}, \bibinfo {author} {\bibfnamefont
  {R.}~\bibnamefont {Trivedi}},\ and\ \bibinfo {author} {\bibfnamefont
  {A.}~\bibnamefont {Karma}},\ }\href
  {https://doi.org/10.1016/j.actamat.2018.03.012} {\bibfield  {journal}
  {\bibinfo  {journal} {Acta Materialia}\ }\textbf {\bibinfo {volume} {150}},\
  \bibinfo {pages} {139} (\bibinfo {year} {2018})}\BibitemShut {NoStop}%
\bibitem [{\citenamefont {Ghosh}\ \emph {et~al.}(2018)\citenamefont {Ghosh},
  \citenamefont {Ofori-Opoku},\ and\ \citenamefont
  {Guyer}}]{Ghosh2018SimulationSuperalloys}%
  \BibitemOpen
  \bibfield  {author} {\bibinfo {author} {\bibfnamefont {S.}~\bibnamefont
  {Ghosh}}, \bibinfo {author} {\bibfnamefont {N.}~\bibnamefont {Ofori-Opoku}},\
  and\ \bibinfo {author} {\bibfnamefont {J.~E.}\ \bibnamefont {Guyer}},\ }\href
  {https://doi.org/10.1016/J.COMMATSCI.2017.12.037} {\bibfield  {journal}
  {\bibinfo  {journal} {Computational Materials Science}\ }\textbf {\bibinfo
  {volume} {144}},\ \bibinfo {pages} {256} (\bibinfo {year}
  {2018})}\BibitemShut {NoStop}%
\bibitem [{\citenamefont {Wang}\ \emph {et~al.}(2020)\citenamefont {Wang},
  \citenamefont {Boussinot}, \citenamefont {H{\"{u}}ter}, \citenamefont
  {Brener},\ and\ \citenamefont {Spatschek}}]{Wang2020ModelingModel}%
  \BibitemOpen
  \bibfield  {author} {\bibinfo {author} {\bibfnamefont {K.}~\bibnamefont
  {Wang}}, \bibinfo {author} {\bibfnamefont {G.}~\bibnamefont {Boussinot}},
  \bibinfo {author} {\bibfnamefont {C.}~\bibnamefont {H{\"{u}}ter}}, \bibinfo
  {author} {\bibfnamefont {E.~A.}\ \bibnamefont {Brener}},\ and\ \bibinfo
  {author} {\bibfnamefont {R.}~\bibnamefont {Spatschek}},\ }\href
  {https://doi.org/10.1103/PHYSREVMATERIALS.4.033802/FIGURES/4/MEDIUM}
  {\bibfield  {journal} {\bibinfo  {journal} {Physical Review Materials}\
  }\textbf {\bibinfo {volume} {4}},\ \bibinfo {pages} {033802} (\bibinfo {year}
  {2020})}\BibitemShut {NoStop}%
\bibitem [{\citenamefont {Kurz}\ and\ \citenamefont
  {Fisher}(1989)}]{kurz1989fundamentals}%
  \BibitemOpen
  \bibfield  {author} {\bibinfo {author} {\bibfnamefont {W.}~\bibnamefont
  {Kurz}}\ and\ \bibinfo {author} {\bibfnamefont {D.~J.}\ \bibnamefont
  {Fisher}},\ }\href@noop {} {\emph {\bibinfo {title} {{Fundamentals of
  solidification}}}}\ (\bibinfo  {publisher} {Trans Tech Publications},\
  \bibinfo {year} {1989})\BibitemShut {NoStop}%
\bibitem [{\citenamefont {Dantzig}\ and\ \citenamefont
  {Rappaz}(2016)}]{Dantzig2016Solidification}%
  \BibitemOpen
  \bibfield  {author} {\bibinfo {author} {\bibfnamefont {J.~A.}\ \bibnamefont
  {Dantzig}}\ and\ \bibinfo {author} {\bibfnamefont {M.}~\bibnamefont
  {Rappaz}},\ }\href@noop {} {\emph {\bibinfo {title} {{Solidification}}}}\
  (\bibinfo  {publisher} {EPFL press},\ \bibinfo {year} {2016})\BibitemShut
  {NoStop}%
\bibitem [{\citenamefont {Pinomaa}\ and\ \citenamefont
  {Provatas}(2019)}]{Pinomaa2019QuantitativeSolidification}%
  \BibitemOpen
  \bibfield  {author} {\bibinfo {author} {\bibfnamefont {T.}~\bibnamefont
  {Pinomaa}}\ and\ \bibinfo {author} {\bibfnamefont {N.}~\bibnamefont
  {Provatas}},\ }\href {https://doi.org/10.1016/J.ACTAMAT.2019.02.009}
  {\bibfield  {journal} {\bibinfo  {journal} {Acta Materialia}\ }\textbf
  {\bibinfo {volume} {168}},\ \bibinfo {pages} {167} (\bibinfo {year}
  {2019})}\BibitemShut {NoStop}%
\bibitem [{\citenamefont {Ahmad}\ \emph {et~al.}(1998)\citenamefont {Ahmad},
  \citenamefont {Wheeler}, \citenamefont {Boettinger},\ and\ \citenamefont
  {McFadden}}]{Ahmad1998SoluteSolidification}%
  \BibitemOpen
  \bibfield  {author} {\bibinfo {author} {\bibfnamefont {N.~A.}\ \bibnamefont
  {Ahmad}}, \bibinfo {author} {\bibfnamefont {A.~A.}\ \bibnamefont {Wheeler}},
  \bibinfo {author} {\bibfnamefont {W.~J.}\ \bibnamefont {Boettinger}},\ and\
  \bibinfo {author} {\bibfnamefont {G.~B.}\ \bibnamefont {McFadden}},\ }\href
  {https://doi.org/10.1103/PhysRevE.58.3436} {\bibfield  {journal} {\bibinfo
  {journal} {Physical Review E}\ }\textbf {\bibinfo {volume} {58}},\ \bibinfo
  {pages} {3436} (\bibinfo {year} {1998})}\BibitemShut {NoStop}%
\bibitem [{\citenamefont {Danilov}\ and\ \citenamefont
  {Nestler}(2006)}]{Danilov2006Phase-fieldAlloy}%
  \BibitemOpen
  \bibfield  {author} {\bibinfo {author} {\bibfnamefont {D.}~\bibnamefont
  {Danilov}}\ and\ \bibinfo {author} {\bibfnamefont {B.}~\bibnamefont
  {Nestler}},\ }\href {https://doi.org/10.1016/J.ACTAMAT.2006.05.045}
  {\bibfield  {journal} {\bibinfo  {journal} {Acta Materialia}\ }\textbf
  {\bibinfo {volume} {54}},\ \bibinfo {pages} {4659} (\bibinfo {year}
  {2006})}\BibitemShut {NoStop}%
\bibitem [{\citenamefont {Steinbach}\ \emph {et~al.}(2012)\citenamefont
  {Steinbach}, \citenamefont {Zhang},\ and\ \citenamefont
  {Plapp}}]{Steinbach2012Phase-fieldDissipation}%
  \BibitemOpen
  \bibfield  {author} {\bibinfo {author} {\bibfnamefont {I.}~\bibnamefont
  {Steinbach}}, \bibinfo {author} {\bibfnamefont {L.}~\bibnamefont {Zhang}},\
  and\ \bibinfo {author} {\bibfnamefont {M.}~\bibnamefont {Plapp}},\ }\href
  {https://doi.org/10.1016/J.ACTAMAT.2012.01.035} {\bibfield  {journal}
  {\bibinfo  {journal} {Acta Materialia}\ }\textbf {\bibinfo {volume} {60}},\
  \bibinfo {pages} {2689} (\bibinfo {year} {2012})}\BibitemShut {NoStop}%
\bibitem [{\citenamefont {Kavousi}\ and\ \citenamefont
  {Asle~Zaeem}(2021)}]{Kavousi2021QuantitativeSolidification}%
  \BibitemOpen
  \bibfield  {author} {\bibinfo {author} {\bibfnamefont {S.}~\bibnamefont
  {Kavousi}}\ and\ \bibinfo {author} {\bibfnamefont {M.}~\bibnamefont
  {Asle~Zaeem}},\ }\href {https://doi.org/10.1016/J.ACTAMAT.2020.116562}
  {\bibfield  {journal} {\bibinfo  {journal} {Acta Materialia}\ }\textbf
  {\bibinfo {volume} {205}},\ \bibinfo {pages} {116562} (\bibinfo {year}
  {2021})}\BibitemShut {NoStop}%
\bibitem [{\citenamefont {Boettinger}\ \emph {et~al.}(1984)\citenamefont
  {Boettinger}, \citenamefont {Shechtman}, \citenamefont {Schaefer},\ and\
  \citenamefont {Biancaniello}}]{Boettinger1984TheAlloys}%
  \BibitemOpen
  \bibfield  {author} {\bibinfo {author} {\bibfnamefont {W.~J.}\ \bibnamefont
  {Boettinger}}, \bibinfo {author} {\bibfnamefont {D.}~\bibnamefont
  {Shechtman}}, \bibinfo {author} {\bibfnamefont {R.~J.}\ \bibnamefont
  {Schaefer}},\ and\ \bibinfo {author} {\bibfnamefont {F.~S.}\ \bibnamefont
  {Biancaniello}},\ }\href {https://doi.org/10.1007/BF02644387} {\bibfield
  {journal} {\bibinfo  {journal} {Metallurgical Transactions A 1984 15:1}\
  }\textbf {\bibinfo {volume} {15}},\ \bibinfo {pages} {55} (\bibinfo {year}
  {1984})}\BibitemShut {NoStop}%
\bibitem [{\citenamefont {Zimmermann}\ \emph {et~al.}(1991)\citenamefont
  {Zimmermann}, \citenamefont {Carrard}, \citenamefont {Gremaud},\ and\
  \citenamefont {Kurz}}]{Zimmermann1991CharacterizationAlloys}%
  \BibitemOpen
  \bibfield  {author} {\bibinfo {author} {\bibfnamefont {M.}~\bibnamefont
  {Zimmermann}}, \bibinfo {author} {\bibfnamefont {M.}~\bibnamefont {Carrard}},
  \bibinfo {author} {\bibfnamefont {M.}~\bibnamefont {Gremaud}},\ and\ \bibinfo
  {author} {\bibfnamefont {W.}~\bibnamefont {Kurz}},\ }\href
  {https://doi.org/10.1016/0921-5093(91)90973-Q} {\bibfield  {journal}
  {\bibinfo  {journal} {Materials Science and Engineering: A}\ }\textbf
  {\bibinfo {volume} {134}},\ \bibinfo {pages} {1278} (\bibinfo {year}
  {1991})}\BibitemShut {NoStop}%
\bibitem [{\citenamefont {Gremaud}\ \emph {et~al.}(1991)\citenamefont
  {Gremaud}, \citenamefont {Carrard},\ and\ \citenamefont
  {Kurz}}]{Gremaud1991BandingTreatment}%
  \BibitemOpen
  \bibfield  {author} {\bibinfo {author} {\bibfnamefont {M.}~\bibnamefont
  {Gremaud}}, \bibinfo {author} {\bibfnamefont {M.}~\bibnamefont {Carrard}},\
  and\ \bibinfo {author} {\bibfnamefont {W.}~\bibnamefont {Kurz}},\ }\href
  {https://doi.org/10.1016/0956-7151(91)90228-S} {\bibfield  {journal}
  {\bibinfo  {journal} {Acta Metallurgica et Materialia}\ }\textbf {\bibinfo
  {volume} {39}},\ \bibinfo {pages} {1431} (\bibinfo {year}
  {1991})}\BibitemShut {NoStop}%
\bibitem [{\citenamefont {Carrard}\ \emph {et~al.}(1992)\citenamefont
  {Carrard}, \citenamefont {Gremaud}, \citenamefont {Zimmermann},\ and\
  \citenamefont {Kurz}}]{Carrard1992AboutAlloys}%
  \BibitemOpen
  \bibfield  {author} {\bibinfo {author} {\bibfnamefont {M.}~\bibnamefont
  {Carrard}}, \bibinfo {author} {\bibfnamefont {M.}~\bibnamefont {Gremaud}},
  \bibinfo {author} {\bibfnamefont {M.}~\bibnamefont {Zimmermann}},\ and\
  \bibinfo {author} {\bibfnamefont {W.}~\bibnamefont {Kurz}},\ }\href
  {https://doi.org/10.1016/0956-7151(92)90076-Q} {\bibfield  {journal}
  {\bibinfo  {journal} {Acta Metallurgica et Materialia}\ }\textbf {\bibinfo
  {volume} {40}},\ \bibinfo {pages} {983} (\bibinfo {year} {1992})}\BibitemShut
  {NoStop}%
\bibitem [{\citenamefont {Gill}\ and\ \citenamefont
  {Kurz}(1993)}]{Gill1993RapidlyMap}%
  \BibitemOpen
  \bibfield  {author} {\bibinfo {author} {\bibfnamefont {S.~C.}\ \bibnamefont
  {Gill}}\ and\ \bibinfo {author} {\bibfnamefont {W.}~\bibnamefont {Kurz}},\
  }\href {https://doi.org/10.1016/0956-7151(93)90237-M} {\bibfield  {journal}
  {\bibinfo  {journal} {Acta Metallurgica et Materialia}\ }\textbf {\bibinfo
  {volume} {41}},\ \bibinfo {pages} {3563} (\bibinfo {year}
  {1993})}\BibitemShut {NoStop}%
\bibitem [{\citenamefont {Gill}\ and\ \citenamefont
  {Kurz}(1995)}]{Gill1995RapidlyMap}%
  \BibitemOpen
  \bibfield  {author} {\bibinfo {author} {\bibfnamefont {S.~C.}\ \bibnamefont
  {Gill}}\ and\ \bibinfo {author} {\bibfnamefont {W.}~\bibnamefont {Kurz}},\
  }\href {https://doi.org/10.1016/0956-7151(95)90269-4} {\bibfield  {journal}
  {\bibinfo  {journal} {Acta Metallurgica et Materialia}\ }\textbf {\bibinfo
  {volume} {43}},\ \bibinfo {pages} {139} (\bibinfo {year} {1995})}\BibitemShut
  {NoStop}%
\bibitem [{\citenamefont {Gremaud}\ \emph {et~al.}(1990)\citenamefont
  {Gremaud}, \citenamefont {Carrard},\ and\ \citenamefont
  {Kurz}}]{Gremaud1990TheTreatment}%
  \BibitemOpen
  \bibfield  {author} {\bibinfo {author} {\bibfnamefont {M.}~\bibnamefont
  {Gremaud}}, \bibinfo {author} {\bibfnamefont {M.}~\bibnamefont {Carrard}},\
  and\ \bibinfo {author} {\bibfnamefont {W.}~\bibnamefont {Kurz}},\ }\href
  {https://doi.org/10.1016/0956-7151(90)90271-H} {\bibfield  {journal}
  {\bibinfo  {journal} {Acta Metallurgica et Materialia}\ }\textbf {\bibinfo
  {volume} {38}},\ \bibinfo {pages} {2587} (\bibinfo {year}
  {1990})}\BibitemShut {NoStop}%
\bibitem [{\citenamefont {Kurz}\ and\ \citenamefont
  {Trivedi}(1996)}]{Kurz1996BandedMicrostructures}%
  \BibitemOpen
  \bibfield  {author} {\bibinfo {author} {\bibfnamefont {W.}~\bibnamefont
  {Kurz}}\ and\ \bibinfo {author} {\bibfnamefont {R.}~\bibnamefont {Trivedi}},\
  }\href {https://doi.org/10.1007/BF02648951} {\bibfield  {journal} {\bibinfo
  {journal} {Metallurgical and Materials Transactions A 1996 27:3}\ }\textbf
  {\bibinfo {volume} {27}},\ \bibinfo {pages} {625} (\bibinfo {year}
  {1996})}\BibitemShut {NoStop}%
\bibitem [{\citenamefont {McKeown}\ \emph {et~al.}(2014)\citenamefont
  {McKeown}, \citenamefont {Kulovits}, \citenamefont {Liu}, \citenamefont
  {Zweiacker}, \citenamefont {Reed}, \citenamefont {Lagrange}, \citenamefont
  {Wiezorek},\ and\ \citenamefont {Campbell}}]{McKeown2014InAlloy}%
  \BibitemOpen
  \bibfield  {author} {\bibinfo {author} {\bibfnamefont {J.~T.}\ \bibnamefont
  {McKeown}}, \bibinfo {author} {\bibfnamefont {A.~K.}\ \bibnamefont
  {Kulovits}}, \bibinfo {author} {\bibfnamefont {C.}~\bibnamefont {Liu}},
  \bibinfo {author} {\bibfnamefont {K.}~\bibnamefont {Zweiacker}}, \bibinfo
  {author} {\bibfnamefont {B.~W.}\ \bibnamefont {Reed}}, \bibinfo {author}
  {\bibfnamefont {T.}~\bibnamefont {Lagrange}}, \bibinfo {author}
  {\bibfnamefont {J.~M.}\ \bibnamefont {Wiezorek}},\ and\ \bibinfo {author}
  {\bibfnamefont {G.~H.}\ \bibnamefont {Campbell}},\ }\href
  {https://doi.org/10.1016/J.ACTAMAT.2013.11.046} {\bibfield  {journal}
  {\bibinfo  {journal} {Acta Materialia}\ }\textbf {\bibinfo {volume} {65}},\
  \bibinfo {pages} {56} (\bibinfo {year} {2014})}\BibitemShut {NoStop}%
\bibitem [{\citenamefont {McKeown}\ \emph {et~al.}(2016)\citenamefont
  {McKeown}, \citenamefont {Zweiacker}, \citenamefont {Liu}, \citenamefont
  {Coughlin}, \citenamefont {Clarke}, \citenamefont {Baldwin}, \citenamefont
  {Gibbs}, \citenamefont {Roehling}, \citenamefont {Imhoff}, \citenamefont
  {Gibbs}, \citenamefont {Tourret}, \citenamefont {Wiezorek},\ and\
  \citenamefont {Campbell}}]{McKeown2016Time-ResolvedManufacturing}%
  \BibitemOpen
  \bibfield  {author} {\bibinfo {author} {\bibfnamefont {J.~T.}\ \bibnamefont
  {McKeown}}, \bibinfo {author} {\bibfnamefont {K.}~\bibnamefont {Zweiacker}},
  \bibinfo {author} {\bibfnamefont {C.}~\bibnamefont {Liu}}, \bibinfo {author}
  {\bibfnamefont {D.~R.}\ \bibnamefont {Coughlin}}, \bibinfo {author}
  {\bibfnamefont {A.~J.}\ \bibnamefont {Clarke}}, \bibinfo {author}
  {\bibfnamefont {J.~K.}\ \bibnamefont {Baldwin}}, \bibinfo {author}
  {\bibfnamefont {J.~W.}\ \bibnamefont {Gibbs}}, \bibinfo {author}
  {\bibfnamefont {J.~D.}\ \bibnamefont {Roehling}}, \bibinfo {author}
  {\bibfnamefont {S.~D.}\ \bibnamefont {Imhoff}}, \bibinfo {author}
  {\bibfnamefont {P.~J.}\ \bibnamefont {Gibbs}}, \bibinfo {author}
  {\bibfnamefont {D.}~\bibnamefont {Tourret}}, \bibinfo {author} {\bibfnamefont
  {J.~M.}\ \bibnamefont {Wiezorek}},\ and\ \bibinfo {author} {\bibfnamefont
  {G.~H.}\ \bibnamefont {Campbell}},\ }\href
  {https://doi.org/10.1007/S11837-015-1793-X/TABLES/2} {\bibfield  {journal}
  {\bibinfo  {journal} {JOM}\ }\textbf {\bibinfo {volume} {68}},\ \bibinfo
  {pages} {985} (\bibinfo {year} {2016})}\BibitemShut {NoStop}%
\bibitem [{See()}]{SeeSupplemental}%
  \BibitemOpen
  \href@noop {} {\bibinfo {title} {{See Supplemental Material at [URL will be
  inserted by publisher] for parameters, movies, variational formulation,
  continuous growth model and asymptotic analyses, and the measurement of tip
  supersaturation, which includes Refs. [50-57].}}}\BibitemShut {Stop}%
\bibitem [{\citenamefont {Karma}(2003)}]{Karma2003Phase-fieldFormation}%
  \BibitemOpen
  \bibfield  {author} {\bibinfo {author} {\bibfnamefont {A.}~\bibnamefont
  {Karma}},\ }in\ \href {https://link.springer.com/book/9781402013676} {\emph
  {\bibinfo {booktitle} {Thermodynamics, Microstructures and Plasticity}}},\
  \bibinfo {editor} {edited by\ \bibinfo {editor} {\bibfnamefont
  {A.}~\bibnamefont {Finel}}, \bibinfo {editor} {\bibfnamefont
  {D.}~\bibnamefont {Mazi{\`{e}}re}},\ and\ \bibinfo {editor} {\bibfnamefont
  {M.}~\bibnamefont {Veron}}}\ (\bibinfo  {publisher} {Springer},\ \bibinfo
  {address} {Dordrecht},\ \bibinfo {year} {2003})\ pp.\ \bibinfo {pages}
  {65--89}\BibitemShut {NoStop}%
\bibitem [{\citenamefont {Galenko}\ \emph {et~al.}(2011)\citenamefont
  {Galenko}, \citenamefont {Abramova}, \citenamefont {Jou}, \citenamefont
  {Danilov}, \citenamefont {Lebedev},\ and\ \citenamefont
  {Herlach}}]{Galenko2011SoluteStudy}%
  \BibitemOpen
  \bibfield  {author} {\bibinfo {author} {\bibfnamefont {P.~K.}\ \bibnamefont
  {Galenko}}, \bibinfo {author} {\bibfnamefont {E.~V.}\ \bibnamefont
  {Abramova}}, \bibinfo {author} {\bibfnamefont {D.}~\bibnamefont {Jou}},
  \bibinfo {author} {\bibfnamefont {D.~A.}\ \bibnamefont {Danilov}}, \bibinfo
  {author} {\bibfnamefont {V.~G.}\ \bibnamefont {Lebedev}},\ and\ \bibinfo
  {author} {\bibfnamefont {D.~M.}\ \bibnamefont {Herlach}},\ }\href
  {https://doi.org/10.1103/PHYSREVE.84.041143/FIGURES/7/MEDIUM} {\bibfield
  {journal} {\bibinfo  {journal} {Physical Review E - Statistical, Nonlinear,
  and Soft Matter Physics}\ }\textbf {\bibinfo {volume} {84}},\ \bibinfo
  {pages} {041143} (\bibinfo {year} {2011})}\BibitemShut {NoStop}%
\bibitem [{\citenamefont {Pinomaa}\ \emph {et~al.}(2020)\citenamefont
  {Pinomaa}, \citenamefont {McKeown}, \citenamefont {Wiezorek}, \citenamefont
  {Provatas}, \citenamefont {Laukkanen},\ and\ \citenamefont
  {Suhonen}}]{Pinomaa2020PhaseFilms}%
  \BibitemOpen
  \bibfield  {author} {\bibinfo {author} {\bibfnamefont {T.}~\bibnamefont
  {Pinomaa}}, \bibinfo {author} {\bibfnamefont {J.~M.}\ \bibnamefont
  {McKeown}}, \bibinfo {author} {\bibfnamefont {J.~M.}\ \bibnamefont
  {Wiezorek}}, \bibinfo {author} {\bibfnamefont {N.}~\bibnamefont {Provatas}},
  \bibinfo {author} {\bibfnamefont {A.}~\bibnamefont {Laukkanen}},\ and\
  \bibinfo {author} {\bibfnamefont {T.}~\bibnamefont {Suhonen}},\ }\href
  {https://doi.org/10.1016/J.JCRYSGRO.2019.125418} {\bibfield  {journal}
  {\bibinfo  {journal} {Journal of Crystal Growth}\ }\textbf {\bibinfo {volume}
  {532}},\ \bibinfo {pages} {125418} (\bibinfo {year} {2020})}\BibitemShut
  {NoStop}%
\bibitem [{\citenamefont {Mullins}\ and\ \citenamefont
  {Sekerka}(1964)}]{Mullins1964StabilityAlloy}%
  \BibitemOpen
  \bibfield  {author} {\bibinfo {author} {\bibfnamefont {W.~W.}\ \bibnamefont
  {Mullins}}\ and\ \bibinfo {author} {\bibfnamefont {R.~F.}\ \bibnamefont
  {Sekerka}},\ }\href {https://doi.org/10.1063/1.1713333} {\bibfield  {journal}
  {\bibinfo  {journal} {Journal of Applied Physics}\ }\textbf {\bibinfo
  {volume} {35}},\ \bibinfo {pages} {444} (\bibinfo {year} {1964})}\BibitemShut
  {NoStop}%
\bibitem [{\citenamefont {Trivedi}\ and\ \citenamefont
  {Kurz}(1986)}]{Trivedi1986MorphologicalConditions}%
  \BibitemOpen
  \bibfield  {author} {\bibinfo {author} {\bibfnamefont {R.}~\bibnamefont
  {Trivedi}}\ and\ \bibinfo {author} {\bibfnamefont {W.}~\bibnamefont {Kurz}},\
  }\href {https://doi.org/10.1016/0001-6160(86)90112-4} {\bibfield  {journal}
  {\bibinfo  {journal} {Acta Metallurgica}\ }\textbf {\bibinfo {volume} {34}},\
  \bibinfo {pages} {1663} (\bibinfo {year} {1986})}\BibitemShut {NoStop}%
\bibitem [{\citenamefont {Ludwig}\ and\ \citenamefont
  {Kurz}(1996)}]{Ludwig1996DirectStability}%
  \BibitemOpen
  \bibfield  {author} {\bibinfo {author} {\bibfnamefont {A.}~\bibnamefont
  {Ludwig}}\ and\ \bibinfo {author} {\bibfnamefont {W.}~\bibnamefont {Kurz}},\
  }\href {https://doi.org/10.1016/1359-6454(95)00448-3} {\bibfield  {journal}
  {\bibinfo  {journal} {Acta Materialia}\ }\textbf {\bibinfo {volume} {44}},\
  \bibinfo {pages} {3643} (\bibinfo {year} {1996})}\BibitemShut {NoStop}%
\bibitem [{\citenamefont {Boettinger}\ and\ \citenamefont
  {A.~Warren}(1999)}]{Boettinger1999SimulationVelocity}%
  \BibitemOpen
  \bibfield  {author} {\bibinfo {author} {\bibfnamefont {W.~J.}\ \bibnamefont
  {Boettinger}}\ and\ \bibinfo {author} {\bibfnamefont {J.}~\bibnamefont
  {A.~Warren}},\ }\href {https://doi.org/10.1016/S0022-0248(98)01063-X}
  {\bibfield  {journal} {\bibinfo  {journal} {Journal of Crystal Growth}\
  }\textbf {\bibinfo {volume} {200}},\ \bibinfo {pages} {583} (\bibinfo {year}
  {1999})}\BibitemShut {NoStop}%
\bibitem [{\citenamefont {Langer}\ and\ \citenamefont
  {Hong}(1986)}]{Langer1986SolvabilityAnisotropy}%
  \BibitemOpen
  \bibfield  {author} {\bibinfo {author} {\bibfnamefont {J.~S.}\ \bibnamefont
  {Langer}}\ and\ \bibinfo {author} {\bibfnamefont {D.~C.}\ \bibnamefont
  {Hong}},\ }\href {https://doi.org/10.1103/PhysRevA.34.1462} {\bibfield
  {journal} {\bibinfo  {journal} {Physical Review A}\ }\textbf {\bibinfo
  {volume} {34}},\ \bibinfo {pages} {1462} (\bibinfo {year}
  {1986})}\BibitemShut {NoStop}%
\bibitem [{\citenamefont {Brener}(1990)}]{Brener1990EffectsDendrites}%
  \BibitemOpen
  \bibfield  {author} {\bibinfo {author} {\bibfnamefont {E.~A.}\ \bibnamefont
  {Brener}},\ }\href {https://doi.org/10.1016/0022-0248(90)90505-F} {\bibfield
  {journal} {\bibinfo  {journal} {Journal of Crystal Growth}\ }\textbf
  {\bibinfo {volume} {99}},\ \bibinfo {pages} {165} (\bibinfo {year}
  {1990})}\BibitemShut {NoStop}%
\bibitem [{\citenamefont {Brener}\ and\ \citenamefont
  {Mel'nikov}(1991)}]{Brener1991PatternGrowth}%
  \BibitemOpen
  \bibfield  {author} {\bibinfo {author} {\bibfnamefont {E.~A.}\ \bibnamefont
  {Brener}}\ and\ \bibinfo {author} {\bibfnamefont {V.~I.}\ \bibnamefont
  {Mel'nikov}},\ }\href {https://doi.org/10.1080/00018739100101472} {\bibfield
  {journal} {\bibinfo  {journal} {Advances in Physics}\ }\textbf {\bibinfo
  {volume} {40}},\ \bibinfo {pages} {53} (\bibinfo {year} {1991})}\BibitemShut
  {NoStop}%
\bibitem [{\citenamefont {Bragard}\ \emph {et~al.}(2002)\citenamefont
  {Bragard}, \citenamefont {Karma}, \citenamefont {Lee},\ and\ \citenamefont
  {Plapp}}]{Bragard2002}%
  \BibitemOpen
  \bibfield  {author} {\bibinfo {author} {\bibfnamefont {J.}~\bibnamefont
  {Bragard}}, \bibinfo {author} {\bibfnamefont {A.}~\bibnamefont {Karma}},
  \bibinfo {author} {\bibfnamefont {Y.~H.}\ \bibnamefont {Lee}},\ and\ \bibinfo
  {author} {\bibfnamefont {M.}~\bibnamefont {Plapp}},\ }\bibfield  {journal}
  {\bibinfo  {journal} {Interface Science}\ }\textbf {\bibinfo {volume} {10}},\
  \href {https://doi.org/10.1023/A:1015815928191} {10.1023/A:1015815928191}
  (\bibinfo {year} {2002})\BibitemShut {NoStop}%
\bibitem [{\citenamefont {Karma}\ and\ \citenamefont
  {Sarkissian}(1992)}]{Karma1992DynamicsSolidification}%
  \BibitemOpen
  \bibfield  {author} {\bibinfo {author} {\bibfnamefont {A.}~\bibnamefont
  {Karma}}\ and\ \bibinfo {author} {\bibfnamefont {A.}~\bibnamefont
  {Sarkissian}},\ }\href {https://doi.org/10.1103/PhysRevLett.68.2616}
  {\bibfield  {journal} {\bibinfo  {journal} {Physical Review Letters}\
  }\textbf {\bibinfo {volume} {68}},\ \bibinfo {pages} {2616} (\bibinfo {year}
  {1992})}\BibitemShut {NoStop}%
\bibitem [{\citenamefont {Karma}\ and\ \citenamefont
  {Sarkissian}(1993)}]{Karma1993InterfaceSolidification}%
  \BibitemOpen
  \bibfield  {author} {\bibinfo {author} {\bibfnamefont {A.}~\bibnamefont
  {Karma}}\ and\ \bibinfo {author} {\bibfnamefont {A.}~\bibnamefont
  {Sarkissian}},\ }\href {https://doi.org/10.1103/PhysRevE.47.513} {\bibfield
  {journal} {\bibinfo  {journal} {Physical Review E}\ }\textbf {\bibinfo
  {volume} {47}},\ \bibinfo {pages} {513} (\bibinfo {year} {1993})}\BibitemShut
  {NoStop}%
\bibitem [{\citenamefont {Lee}\ \emph {et~al.}(2004)\citenamefont {Lee},
  \citenamefont {Liu}, \citenamefont {Miyahara},\ and\ \citenamefont
  {Trivedi}}]{Lee2004Diffusion-coefficientAlloys}%
  \BibitemOpen
  \bibfield  {author} {\bibinfo {author} {\bibfnamefont {J.~H.}\ \bibnamefont
  {Lee}}, \bibinfo {author} {\bibfnamefont {S.}~\bibnamefont {Liu}}, \bibinfo
  {author} {\bibfnamefont {H.}~\bibnamefont {Miyahara}},\ and\ \bibinfo
  {author} {\bibfnamefont {R.}~\bibnamefont {Trivedi}},\ }\href
  {https://doi.org/10.1007/S11663-004-0085-6} {\bibfield  {journal} {\bibinfo
  {journal} {Metallurgical and Materials Transactions B 2004 35:5}\ }\textbf
  {\bibinfo {volume} {35}},\ \bibinfo {pages} {909} (\bibinfo {year}
  {2004})}\BibitemShut {NoStop}%
\bibitem [{\citenamefont {G{\"{u}}nd{\"{u}}z}\ and\ \citenamefont
  {Hunt}(1985)}]{Gunduz1985TheSystems}%
  \BibitemOpen
  \bibfield  {author} {\bibinfo {author} {\bibfnamefont {M.}~\bibnamefont
  {G{\"{u}}nd{\"{u}}z}}\ and\ \bibinfo {author} {\bibfnamefont {J.~D.}\
  \bibnamefont {Hunt}},\ }\href {https://doi.org/10.1016/0001-6160(85)90161-0}
  {\bibfield  {journal} {\bibinfo  {journal} {Acta Metallurgica}\ }\textbf
  {\bibinfo {volume} {33}},\ \bibinfo {pages} {1651} (\bibinfo {year}
  {1985})}\BibitemShut {NoStop}%
\bibitem [{\citenamefont {Mendelev}\ \emph {et~al.}(2010)\citenamefont
  {Mendelev}, \citenamefont {Rahman}, \citenamefont {Hoyt},\ and\ \citenamefont
  {Asta}}]{Mendelev2010Molecular-dynamicsMetals}%
  \BibitemOpen
  \bibfield  {author} {\bibinfo {author} {\bibfnamefont {M.~I.}\ \bibnamefont
  {Mendelev}}, \bibinfo {author} {\bibfnamefont {M.~J.}\ \bibnamefont
  {Rahman}}, \bibinfo {author} {\bibfnamefont {J.~J.}\ \bibnamefont {Hoyt}},\
  and\ \bibinfo {author} {\bibfnamefont {M.}~\bibnamefont {Asta}},\ }\href
  {https://doi.org/10.1088/0965-0393/18/7/074002} {\bibfield  {journal}
  {\bibinfo  {journal} {Modelling and Simulation in Materials Science and
  Engineering}\ }\textbf {\bibinfo {volume} {18}},\ \bibinfo {pages} {074002}
  (\bibinfo {year} {2010})}\BibitemShut {NoStop}%
\bibitem [{\citenamefont {Ji}\ \emph {et~al.}(2022)\citenamefont {Ji},
  \citenamefont {Tabrizi},\ and\ \citenamefont
  {Karma}}]{Ji2022IsotropicSolidification}%
  \BibitemOpen
  \bibfield  {author} {\bibinfo {author} {\bibfnamefont {K.}~\bibnamefont
  {Ji}}, \bibinfo {author} {\bibfnamefont {A.~M.}\ \bibnamefont {Tabrizi}},\
  and\ \bibinfo {author} {\bibfnamefont {A.}~\bibnamefont {Karma}},\ }\href
  {https://doi.org/10.1016/J.JCP.2022.111069} {\bibfield  {journal} {\bibinfo
  {journal} {Journal of Computational Physics}\ }\textbf {\bibinfo {volume}
  {457}},\ \bibinfo {pages} {111069} (\bibinfo {year} {2022})}\BibitemShut
  {NoStop}%
\bibitem [{\citenamefont {Aziz}(1982)}]{Aziz1982ModelSolidification}%
  \BibitemOpen
  \bibfield  {author} {\bibinfo {author} {\bibfnamefont {M.~J.}\ \bibnamefont
  {Aziz}},\ }\href {https://doi.org/10.1063/1.329867} {\bibfield  {journal}
  {\bibinfo  {journal} {Journal of Applied Physics}\ }\textbf {\bibinfo
  {volume} {53}},\ \bibinfo {pages} {1158} (\bibinfo {year}
  {1982})}\BibitemShut {NoStop}%
\bibitem [{\citenamefont {Aziz}\ and\ \citenamefont
  {Kaplan}(1988)}]{Aziz1988ContinuousSolidification}%
  \BibitemOpen
  \bibfield  {author} {\bibinfo {author} {\bibfnamefont {M.~J.}\ \bibnamefont
  {Aziz}}\ and\ \bibinfo {author} {\bibfnamefont {T.}~\bibnamefont {Kaplan}},\
  }\href {https://doi.org/10.1016/0001-6160(88)90333-1} {\bibfield  {journal}
  {\bibinfo  {journal} {Acta Metallurgica}\ }\textbf {\bibinfo {volume} {36}},\
  \bibinfo {pages} {2335} (\bibinfo {year} {1988})}\BibitemShut {NoStop}%
\bibitem [{\citenamefont {Aziz}\ and\ \citenamefont
  {Boettinger}(1994)}]{Aziz1994OnSolidification}%
  \BibitemOpen
  \bibfield  {author} {\bibinfo {author} {\bibfnamefont {M.~J.}\ \bibnamefont
  {Aziz}}\ and\ \bibinfo {author} {\bibfnamefont {W.~J.}\ \bibnamefont
  {Boettinger}},\ }\href {https://doi.org/10.1016/0956-7151(94)90507-X}
  {\bibfield  {journal} {\bibinfo  {journal} {Acta Metallurgica et Materialia}\
  }\textbf {\bibinfo {volume} {42}},\ \bibinfo {pages} {527} (\bibinfo {year}
  {1994})}\BibitemShut {NoStop}%
\bibitem [{\citenamefont {Ivantsov}(1947)}]{Ivantsov1947TemperatureMelt}%
  \BibitemOpen
  \bibfield  {author} {\bibinfo {author} {\bibfnamefont {G.~P.}\ \bibnamefont
  {Ivantsov}},\ }\href@noop {} {\bibfield  {journal} {\bibinfo  {journal}
  {Doklady Akademii Nauk, SSSR}\ }\textbf {\bibinfo {volume} {58}} (\bibinfo
  {year} {1947})}\BibitemShut {NoStop}%
\end{thebibliography}%


\begin{thebibliography}{15}%
\makeatletter
\providecommand \@ifxundefined [1]{%
 \@ifx{#1\undefined}
}%
\providecommand \@ifnum [1]{%
 \ifnum #1\expandafter \@firstoftwo
 \else \expandafter \@secondoftwo
 \fi
}%
\providecommand \@ifx [1]{%
 \ifx #1\expandafter \@firstoftwo
 \else \expandafter \@secondoftwo
 \fi
}%
\providecommand \natexlab [1]{#1}%
\providecommand \enquote  [1]{``#1''}%
\providecommand \bibnamefont  [1]{#1}%
\providecommand \bibfnamefont [1]{#1}%
\providecommand \citenamefont [1]{#1}%
\providecommand \href@noop [0]{\@secondoftwo}%
\providecommand \href [0]{\begingroup \@sanitize@url \@href}%
\providecommand \@href[1]{\@@startlink{#1}\@@href}%
\providecommand \@@href[1]{\endgroup#1\@@endlink}%
\providecommand \@sanitize@url [0]{\catcode `\\12\catcode `\$12\catcode
  `\&12\catcode `\#12\catcode `\^12\catcode `\_12\catcode `\%12\relax}%
\providecommand \@@startlink[1]{}%
\providecommand \@@endlink[0]{}%
\providecommand \url  [0]{\begingroup\@sanitize@url \@url }%
\providecommand \@url [1]{\endgroup\@href {#1}{\urlprefix }}%
\providecommand \urlprefix  [0]{URL }%
\providecommand \Eprint [0]{\href }%
\providecommand \doibase [0]{https://doi.org/}%
\providecommand \selectlanguage [0]{\@gobble}%
\providecommand \bibinfo  [0]{\@secondoftwo}%
\providecommand \bibfield  [0]{\@secondoftwo}%
\providecommand \translation [1]{[#1]}%
\providecommand \BibitemOpen [0]{}%
\providecommand \bibitemStop [0]{}%
\providecommand \bibitemNoStop [0]{.\EOS\space}%
\providecommand \EOS [0]{\spacefactor3000\relax}%
\providecommand \BibitemShut  [1]{\csname bibitem#1\endcsname}%
\let\auto@bib@innerbib\@empty
\bibitem [{\citenamefont {Lee}\ \emph {et~al.}(2004)\citenamefont {Lee},
  \citenamefont {Liu}, \citenamefont {Miyahara},\ and\ \citenamefont
  {Trivedi}}]{Lee2004Diffusion-coefficientAlloys}%
  \BibitemOpen
  \bibfield  {author} {\bibinfo {author} {\bibfnamefont {J.~H.}\ \bibnamefont
  {Lee}}, \bibinfo {author} {\bibfnamefont {S.}~\bibnamefont {Liu}}, \bibinfo
  {author} {\bibfnamefont {H.}~\bibnamefont {Miyahara}},\ and\ \bibinfo
  {author} {\bibfnamefont {R.}~\bibnamefont {Trivedi}},\ }\href
  {https://doi.org/10.1007/S11663-004-0085-6} {\bibfield  {journal} {\bibinfo
  {journal} {Metallurgical and Materials Transactions B 2004 35:5}\ }\textbf
  {\bibinfo {volume} {35}},\ \bibinfo {pages} {909} (\bibinfo {year}
  {2004})}\BibitemShut {NoStop}%
\bibitem [{\citenamefont {Kurz}\ and\ \citenamefont
  {Fisher}(1989)}]{kurz1989fundamentals}%
  \BibitemOpen
  \bibfield  {author} {\bibinfo {author} {\bibfnamefont {W.}~\bibnamefont
  {Kurz}}\ and\ \bibinfo {author} {\bibfnamefont {D.~J.}\ \bibnamefont
  {Fisher}},\ }\href@noop {} {\emph {\bibinfo {title} {{Fundamentals of
  solidification}}}}\ (\bibinfo  {publisher} {Trans Tech Publications},\
  \bibinfo {year} {1989})\BibitemShut {NoStop}%
\bibitem [{\citenamefont {Clarke}\ \emph {et~al.}(2017)\citenamefont {Clarke},
  \citenamefont {Tourret}, \citenamefont {Song}, \citenamefont {Imhoff},
  \citenamefont {Gibbs}, \citenamefont {Gibbs}, \citenamefont {Fezzaa},\ and\
  \citenamefont {Karma}}]{Clarke2017}%
  \BibitemOpen
  \bibfield  {author} {\bibinfo {author} {\bibfnamefont {A.~J.}\ \bibnamefont
  {Clarke}}, \bibinfo {author} {\bibfnamefont {D.}~\bibnamefont {Tourret}},
  \bibinfo {author} {\bibfnamefont {Y.}~\bibnamefont {Song}}, \bibinfo {author}
  {\bibfnamefont {S.~D.}\ \bibnamefont {Imhoff}}, \bibinfo {author}
  {\bibfnamefont {P.~J.}\ \bibnamefont {Gibbs}}, \bibinfo {author}
  {\bibfnamefont {J.~W.}\ \bibnamefont {Gibbs}}, \bibinfo {author}
  {\bibfnamefont {K.}~\bibnamefont {Fezzaa}},\ and\ \bibinfo {author}
  {\bibfnamefont {A.}~\bibnamefont {Karma}},\ }\bibfield  {journal} {\bibinfo
  {journal} {Acta Materialia}\ }\href
  {https://doi.org/10.1016/j.actamat.2017.02.047}
  {10.1016/j.actamat.2017.02.047} (\bibinfo {year} {2017})\BibitemShut
  {NoStop}%
\bibitem [{\citenamefont {G{\"{u}}nd{\"{u}}z}\ and\ \citenamefont
  {Hunt}(1985)}]{Gunduz1985TheSystems}%
  \BibitemOpen
  \bibfield  {author} {\bibinfo {author} {\bibfnamefont {M.}~\bibnamefont
  {G{\"{u}}nd{\"{u}}z}}\ and\ \bibinfo {author} {\bibfnamefont {J.~D.}\
  \bibnamefont {Hunt}},\ }\href {https://doi.org/10.1016/0001-6160(85)90161-0}
  {\bibfield  {journal} {\bibinfo  {journal} {Acta Metallurgica}\ }\textbf
  {\bibinfo {volume} {33}},\ \bibinfo {pages} {1651} (\bibinfo {year}
  {1985})}\BibitemShut {NoStop}%
\bibitem [{\citenamefont {Mendelev}\ \emph {et~al.}(2010)\citenamefont
  {Mendelev}, \citenamefont {Rahman}, \citenamefont {Hoyt},\ and\ \citenamefont
  {Asta}}]{Mendelev2010Molecular-dynamicsMetals}%
  \BibitemOpen
  \bibfield  {author} {\bibinfo {author} {\bibfnamefont {M.~I.}\ \bibnamefont
  {Mendelev}}, \bibinfo {author} {\bibfnamefont {M.~J.}\ \bibnamefont
  {Rahman}}, \bibinfo {author} {\bibfnamefont {J.~J.}\ \bibnamefont {Hoyt}},\
  and\ \bibinfo {author} {\bibfnamefont {M.}~\bibnamefont {Asta}},\ }\href
  {https://doi.org/10.1088/0965-0393/18/7/074002} {\bibfield  {journal}
  {\bibinfo  {journal} {Modelling and Simulation in Materials Science and
  Engineering}\ }\textbf {\bibinfo {volume} {18}},\ \bibinfo {pages} {074002}
  (\bibinfo {year} {2010})}\BibitemShut {NoStop}%
\bibitem [{\citenamefont {Carrard}\ \emph {et~al.}(1992)\citenamefont
  {Carrard}, \citenamefont {Gremaud}, \citenamefont {Zimmermann},\ and\
  \citenamefont {Kurz}}]{Carrard1992AboutAlloys}%
  \BibitemOpen
  \bibfield  {author} {\bibinfo {author} {\bibfnamefont {M.}~\bibnamefont
  {Carrard}}, \bibinfo {author} {\bibfnamefont {M.}~\bibnamefont {Gremaud}},
  \bibinfo {author} {\bibfnamefont {M.}~\bibnamefont {Zimmermann}},\ and\
  \bibinfo {author} {\bibfnamefont {W.}~\bibnamefont {Kurz}},\ }\href
  {https://doi.org/10.1016/0956-7151(92)90076-Q} {\bibfield  {journal}
  {\bibinfo  {journal} {Acta Metallurgica et Materialia}\ }\textbf {\bibinfo
  {volume} {40}},\ \bibinfo {pages} {983} (\bibinfo {year} {1992})}\BibitemShut
  {NoStop}%
\bibitem [{\citenamefont {Song}\ \emph {et~al.}(2018)\citenamefont {Song},
  \citenamefont {Tourret}, \citenamefont {Mota}, \citenamefont {Pereda},
  \citenamefont {Billia}, \citenamefont {Bergeon}, \citenamefont {Trivedi},\
  and\ \citenamefont {Karma}}]{Song2018Thermal-fieldSolidification}%
  \BibitemOpen
  \bibfield  {author} {\bibinfo {author} {\bibfnamefont {Y.}~\bibnamefont
  {Song}}, \bibinfo {author} {\bibfnamefont {D.}~\bibnamefont {Tourret}},
  \bibinfo {author} {\bibfnamefont {F.~L.}\ \bibnamefont {Mota}}, \bibinfo
  {author} {\bibfnamefont {J.}~\bibnamefont {Pereda}}, \bibinfo {author}
  {\bibfnamefont {B.}~\bibnamefont {Billia}}, \bibinfo {author} {\bibfnamefont
  {N.}~\bibnamefont {Bergeon}}, \bibinfo {author} {\bibfnamefont
  {R.}~\bibnamefont {Trivedi}},\ and\ \bibinfo {author} {\bibfnamefont
  {A.}~\bibnamefont {Karma}},\ }\href
  {https://doi.org/10.1016/j.actamat.2018.03.012} {\bibfield  {journal}
  {\bibinfo  {journal} {Acta Materialia}\ }\textbf {\bibinfo {volume} {150}},\
  \bibinfo {pages} {139} (\bibinfo {year} {2018})}\BibitemShut {NoStop}%
\bibitem [{\citenamefont {Ji}\ \emph {et~al.}(2022)\citenamefont {Ji},
  \citenamefont {Tabrizi},\ and\ \citenamefont
  {Karma}}]{Ji2022IsotropicSolidification}%
  \BibitemOpen
  \bibfield  {author} {\bibinfo {author} {\bibfnamefont {K.}~\bibnamefont
  {Ji}}, \bibinfo {author} {\bibfnamefont {A.~M.}\ \bibnamefont {Tabrizi}},\
  and\ \bibinfo {author} {\bibfnamefont {A.}~\bibnamefont {Karma}},\ }\href
  {https://doi.org/10.1016/J.JCP.2022.111069} {\bibfield  {journal} {\bibinfo
  {journal} {Journal of Computational Physics}\ }\textbf {\bibinfo {volume}
  {457}},\ \bibinfo {pages} {111069} (\bibinfo {year} {2022})}\BibitemShut
  {NoStop}%
\bibitem [{Note1()}]{Note1}%
  \BibitemOpen
  \bibinfo {note} {Movies will be available in the published
  version.}\BibitemShut {Stop}%
\bibitem [{\citenamefont {Karma}(2003)}]{Karma2003Phase-fieldFormation}%
  \BibitemOpen
  \bibfield  {author} {\bibinfo {author} {\bibfnamefont {A.}~\bibnamefont
  {Karma}},\ }in\ \href {https://link.springer.com/book/9781402013676} {\emph
  {\bibinfo {booktitle} {Thermodynamics, Microstructures and Plasticity}}},\
  \bibinfo {editor} {edited by\ \bibinfo {editor} {\bibfnamefont
  {A.}~\bibnamefont {Finel}}, \bibinfo {editor} {\bibfnamefont
  {D.}~\bibnamefont {Mazi{\`{e}}re}},\ and\ \bibinfo {editor} {\bibfnamefont
  {M.}~\bibnamefont {Veron}}}\ (\bibinfo  {publisher} {Springer},\ \bibinfo
  {address} {Dordrecht},\ \bibinfo {year} {2003})\ pp.\ \bibinfo {pages}
  {65--89}\BibitemShut {NoStop}%
\bibitem [{\citenamefont {Aziz}(1982)}]{Aziz1982ModelSolidification}%
  \BibitemOpen
  \bibfield  {author} {\bibinfo {author} {\bibfnamefont {M.~J.}\ \bibnamefont
  {Aziz}},\ }\href {https://doi.org/10.1063/1.329867} {\bibfield  {journal}
  {\bibinfo  {journal} {Journal of Applied Physics}\ }\textbf {\bibinfo
  {volume} {53}},\ \bibinfo {pages} {1158} (\bibinfo {year}
  {1982})}\BibitemShut {NoStop}%
\bibitem [{\citenamefont {Aziz}\ and\ \citenamefont
  {Kaplan}(1988)}]{Aziz1988ContinuousSolidification}%
  \BibitemOpen
  \bibfield  {author} {\bibinfo {author} {\bibfnamefont {M.~J.}\ \bibnamefont
  {Aziz}}\ and\ \bibinfo {author} {\bibfnamefont {T.}~\bibnamefont {Kaplan}},\
  }\href {https://doi.org/10.1016/0001-6160(88)90333-1} {\bibfield  {journal}
  {\bibinfo  {journal} {Acta Metallurgica}\ }\textbf {\bibinfo {volume} {36}},\
  \bibinfo {pages} {2335} (\bibinfo {year} {1988})}\BibitemShut {NoStop}%
\bibitem [{\citenamefont {Aziz}\ and\ \citenamefont
  {Boettinger}(1994)}]{Aziz1994OnSolidification}%
  \BibitemOpen
  \bibfield  {author} {\bibinfo {author} {\bibfnamefont {M.~J.}\ \bibnamefont
  {Aziz}}\ and\ \bibinfo {author} {\bibfnamefont {W.~J.}\ \bibnamefont
  {Boettinger}},\ }\href {https://doi.org/10.1016/0956-7151(94)90507-X}
  {\bibfield  {journal} {\bibinfo  {journal} {Acta Metallurgica et Materialia}\
  }\textbf {\bibinfo {volume} {42}},\ \bibinfo {pages} {527} (\bibinfo {year}
  {1994})}\BibitemShut {NoStop}%
\bibitem [{\citenamefont {Ahmad}\ \emph {et~al.}(1998)\citenamefont {Ahmad},
  \citenamefont {Wheeler}, \citenamefont {Boettinger},\ and\ \citenamefont
  {McFadden}}]{Ahmad1998SoluteSolidification}%
  \BibitemOpen
  \bibfield  {author} {\bibinfo {author} {\bibfnamefont {N.~A.}\ \bibnamefont
  {Ahmad}}, \bibinfo {author} {\bibfnamefont {A.~A.}\ \bibnamefont {Wheeler}},
  \bibinfo {author} {\bibfnamefont {W.~J.}\ \bibnamefont {Boettinger}},\ and\
  \bibinfo {author} {\bibfnamefont {G.~B.}\ \bibnamefont {McFadden}},\ }\href
  {https://doi.org/10.1103/PhysRevE.58.3436} {\bibfield  {journal} {\bibinfo
  {journal} {Physical Review E}\ }\textbf {\bibinfo {volume} {58}},\ \bibinfo
  {pages} {3436} (\bibinfo {year} {1998})}\BibitemShut {NoStop}%
\bibitem [{\citenamefont {Ivantsov}(1947)}]{Ivantsov1947TemperatureMelt}%
  \BibitemOpen
  \bibfield  {author} {\bibinfo {author} {\bibfnamefont {G.~P.}\ \bibnamefont
  {Ivantsov}},\ }\href@noop {} {\bibfield  {journal} {\bibinfo  {journal}
  {Doklady Akademii Nauk, SSSR}\ }\textbf {\bibinfo {volume} {58}} (\bibinfo
  {year} {1947})}\BibitemShut {NoStop}%
\end{thebibliography}%


%

\end{document}


\preprint{APS/123-QED}

\title{Supplemental Material for: Microstructural Pattern Formation during Far-from-Equilibrium Alloy Solidification}

\author{Kaihua Ji}
\affiliation{%
Physics Department and Center for Interdisciplinary Research on Complex Systems, Northeastern University, Boston, Massachusetts 02115, USA
}%
\author{Elaheh Dorari}
\affiliation{%
Physics Department and Center for Interdisciplinary Research on Complex Systems, Northeastern University, Boston, Massachusetts 02115, USA
}%
\author{Amy J. Clarke}
\affiliation{%
Department of Material and Metallurgical Engineering, Colorado School of Mines, Golden, CO, 80401, USA
}%
\author{Alain Karma}%
 \email{a.karma@northeastern.edu}
\affiliation{%
Physics Department and Center for Interdisciplinary Research on Complex Systems, Northeastern University, Boston, Massachusetts 02115, USA
}%


\begin{abstract}
\end{abstract}

\maketitle

\onecolumngrid

\section{Parameters and numerical simulations}

\begin{table}[hbt!]
\caption{\label{tab:table1}
Materials, process, and simulation parameters.
}
\begin{ruledtabular}
\begin{tabular}{lllll}
Symbol & Description & Value & Unit & Ref.\\
\colrule
$c_{\infty}$& Nominal composition & 3, 9 & wt.\% Cu \\
$D_l$ & Solute diffusivity in liquid & 2400 & $\mathrm{\mu m^2 \, s^{-1}}$ & \cite{Lee2004Diffusion-coefficientAlloys} \\
$k_e$ & Equilibrium partition coefficient & 0.14 & & \cite{kurz1989fundamentals}\\
$m_e$ & Equilibrium liquidus slope & $2.6$ & K wt.\%$^{-1}$ Cu& \cite{Clarke2017}\\
$\Gamma$ & Gibbs-Thomson coefficient & 0.24 & $\mathrm{\mu m \, K}$ & \cite{kurz1989fundamentals,Gunduz1985TheSystems} \\
$W_0$ & Atomic interface thickness & 1 & nm\\
$V_d^0$ & Diffusive speed $D_l/W_0$ & 2.4 & $\mathrm{m \, s^{-1}}$ & \\
$\mu_k^0$ & Interface kinetic coefficient & 0.5 & $\mathrm{m \, s^{-1} \, K^{-1}}$ & \cite{Mendelev2010Molecular-dynamicsMetals} \\
$\epsilon_k$ & Kinetic anisotropy strength & 0.1 \\
$\epsilon_s$ & Interface free-energy anisotropy strength & 0.012\\
\\
$G$ & Temperature gradient & $5 \times 10^6$ & K m$^{-1}$ & \cite{Carrard1992AboutAlloys} \\
\\
$W$ & Interface thickness & 1, 3, 5 & $W_0$\\
$\Delta x$ & Grid spacing & 0.8 & $W$ \\
\end{tabular}
\end{ruledtabular}
\end{table}

Phase-field (PF) simulations are performed for directional solidification of Al-Cu alloys with parameters given in Table \ref{tab:table1}. For PF simulations with latent-heat rejection at the solid-liquid interface \cite{Song2018Thermal-fieldSolidification}, thermal properties of the Al-Cu alloy used in the simulation include the thermal diffusivity $5.35 \times 10^{-5} \, \mathrm{m^{2}/s}$, and $L/c_p=$ 340.5 K, where $L$ is the latent heat of fusion per unit volume and $c_p$ is the heat capacity. We implement the PF model on massively parallel graphic processing units (GPU) with the computer unified device architecture (CUDA) programming language. The model equations are solved on a square lattice using a finite difference implementation of spatial derivatives and an Euler explicit time stepping scheme. For leading differential terms that include Laplacian and divergence, we use isotropic discretizations as described in Ref.\ \cite{Ji2022IsotropicSolidification}. The explicit time step is chosen as $\Delta t/\tau_0 = R_t (\Delta x/W)^2/4$, where the coefficient $R_t$ is set to be 0.6 in all simulations. This choice of $\Delta t$ ensures the numerical stability of both evolution equations of the PF and the concentration field for $D_l \tau_0/W^2 \le 1$ ($D_l \tau_0/W^2=0.02$ with the parameters given in Table \ref{tab:table1}). For most simulations, the solid-liquid interface is initially planar and located at the liquidus temperature at rest. A random perturbation $\eta \beta(y)$ to the perpendicular interface location (along $x$-axis) is given to the initial planar interface, where $\eta=0.5\Delta x$ is the noise amplitude and $\beta(y)$ (a function of $y$ along the horizontal interface location) is a random number generated with a flat distribution in the range $[-0.5,\,0.5]$. The initial PF profile is the 1D approximate solution $\phi_0(x)$, and the initial concentration profile is the corresponding equilibrium profile $c(x)=c_{\infty}\exp[b(g(\phi_0(x))+1)]$ at the liquidus temperature. To construct the red curve (steady-state dendrite growth) in Fig.\ 4(e) of the main text, a simulation was first performed at a very low velocity below the onset of banding to form a steady-state dendrite array structure. The steady-state branch of dendrite array solutions was then followed by increasing the velocity in small steps until the onset of the burgeoning instability described in Fig.\ 3 of the main text. All the other simulations corresponding to the banding cycles shown in Figs.\ 4 and 5 of the main text were carried out starting from a stationary interface at the liquidus temperature. 

We compare the efficiency of 2D PF simulations with different $S$ in Table \ref{tab:table2}, where simulations are performed for dendritic array growth for Al-3wt.\% Cu, $V_p=0.12$ m/s, and $G=5\times 10^6$ K/m; the primary spacing $\Lambda = 0.65$ $\mathrm{\mu m}$, the simulation domain size is $4.8\times0.65$ $\mathrm{\mu m^2}$, and the simulated physical time is $10^{-4}$ s. Fig.\ 2(a) of the main text shows cropped results of these simulations. The results demonstrate that different $S$ yield nearly identical shapes with a three orders of magnitude reduction in computation time for $S=5$ compared to $S=1$.

\begin{table}[htbp!]
\caption{\label{tab:table2}
Simulation time and the coefficient $A$ in the interpolation function $q(\phi)=A(1-\phi)/2-(A-1)(1-\phi)^2/4$ for different $S$. Details about the selection of $A$ for different $S$ are described in Subsection \ref{Select_A}.
}
\begin{ruledtabular}
\begin{tabular}{ccc}
$S$ & $A$ & Simulation time\footnote{Time in minutes for PF simulations performed on a Nvidia A100 Graphics Processing Unit.}\\
\colrule
1 & 1 & 1894 \\
3 & 6 & 25 \\
5 & 12 & 4 \\
\end{tabular}
\end{ruledtabular}
\end{table}

\subsection{Determination of the interfacial diffusivity coefficient \textit{A}} \label{Select_A}

The coefficient $A \ge 1$ in the interpolation function $q(\phi)=A(1-\phi)/2-(A-1)(1-\phi)^2/4$ is an adjustable parameter for different $S$. A larger value of $A$ corresponds to a larger solute diffusivity in the interface region. To obtain values of $A$ that yield $S$-independent trapping properties, we first compute the reference $k(V)$ and $m(V)$ curves corresponding to $S=1$ and $A=1$. For a given $S>1$, we then compute $k(V)$ and $m(V)$ curves for different $A$ values and find the value of $A$ that minimizes the departure from the reference curves over a large velocity range of interest (from 0.024 to 10.509 m/s). The approximate ($\phi=\phi_0$) solution is used in this procedure. For each value of $A$, we use a sequence of $N$ velocities $\{V_i\,|\,i=0,1,2,\ldots,N-1\}$ to compute for both $k(V_i)$ and $m(V_i)$, where $V_i=1.5^i V_{\mathrm{min}}$, $V_{\mathrm{min}}=0.024$ m/s, and $N=16$. To quantify the departure of the partition coefficient from its reference value at each velocity $V_i$, we define the quantity $\sqrt{(k_i^S-k_i^1)^2}/k_i^1$, where $k_i^1$ is the reference partition coefficient for $S=1$ and $k_i^S$ is the partition coefficient for $S>1$. For a given $A$, we then compute the difference averaged over all velocities defined by
\begin{equation}
\delta_k = \frac{1}{N}  \sum_{i=0}^{N-1} \left( \frac{\sqrt{(k_i^S-k_i^1)^2}}{k_i^1} \right),
\end{equation}
which quantifies the departure from the entire reference $k(V)$ curve. Similarly, we compute the average difference between the measured and the reference liquidus slope
\begin{equation}
\delta_m = \frac{1}{N}  \sum_{i=0}^{N-1} \left( \frac{\sqrt{(m_i^S-m_i^1)^2}}{m_i^1} \right)
\end{equation}
for a given $A$ to quantify the departure from the entire reference $m(V)$ curve.

\begin{figure}[hbt!]
\includegraphics[scale=1.2]{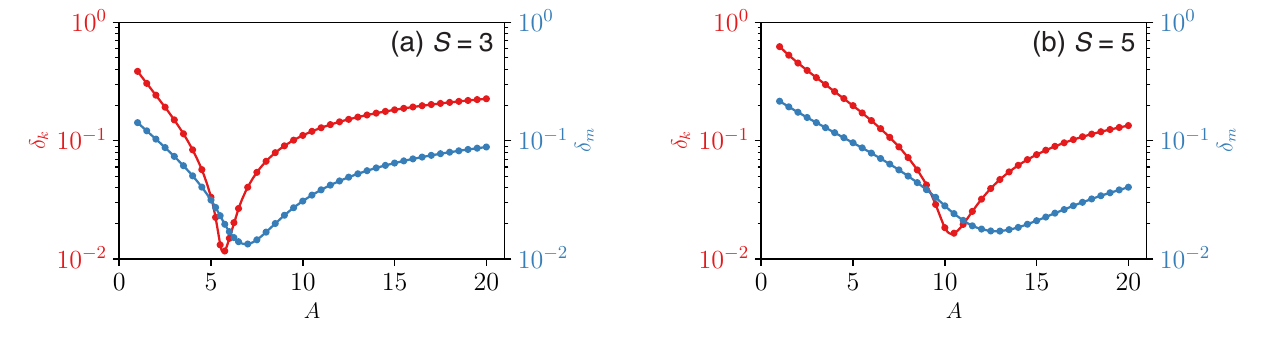}
\caption{\label{fig:figure_km} 
The average difference between the measured ($S>1$) and reference ($S=1$) trapping properties as a function of $A$ for (a) $S=3$ and (b) $S=5$.
}
\end{figure}

As shown in Fig.\ \ref{fig:figure_km}, we compute $\delta_k$ and $\delta_m$ as a function of $A$ using parameters of Al-Cu alloys given in Table \ref{tab:table1}. The $A$ values that minimize $\delta_k$ and $\delta_m$ differ slightly for a given $S$. Therefore, to minimize the departure from both the $k(V)$ and $m(V)$ reference curves, we choose values of $A$ in between the two minima of $\delta_k$ and $\delta_m$ versus $A$. We have checked that the results do not depend sensitively on the choice of the value of $A$ in between those two minima that are relatively close to each other. For concreteness, all the simulations reported in the paper were carried out with $A=6$ and 12 for $S=3$ and 5, respectively. 

\section{Movies}

Movies \footnote{Movies will be available in the published version.} of 2D PF simulations are shown for directional solidification of an Al-3wt.\% Cu alloy with parameters given in Table \ref{tab:table1}. Simulations were performed with $S\equiv W/W_0=5$.

\begin{itemize}
    \item Supplementary\_Movie\_1.mp4: Tip ``burgeoning'' instability during the transition from a steady-state dendrite to planar interface in a PF simulation with $V_p$ approaching a critical velocity 0.88 m/s, where the contours correspond to different PF values, including $\phi=-0.8$, $\phi=-0.4$, $\phi=0.0$ (the solid-liquid interface), $\phi=0.4$, and $\phi=0.8$. The movie corresponds to Fig.\ 3(a) in the main text, but is shown in a frame moving with the solidification front.
    \item Supplementary\_Movie\_2.mp4: Banding cycles in a PF simulation with isotherm velocity 0.46 m/s, where the colormap represents scaled solute concentration $c/c_{\infty}$ and black curves represent solid-liquid interfaces. The movie corresponds to Fig.\ 4(a) in the main text.
    \item Supplementary\_Movie\_3.mp4: A PF simulation with latent-heat rejection at an isotherm velocity 0.96 m/s that corresponds to Fig.\ 4(d) in the main text. In the movie, the colormap represents scaled solute concentration $c/c_{\infty}$ and black curves represent solid-liquid interfaces.
\end{itemize}

\section{Variational formulation}

The total free-energy of a two-phase system for alloy solidification can be written in the form \cite{Karma2003Phase-fieldFormation}
\begin{equation}
F[\phi, c, T]=\int_{d V}\left[\frac{\sigma}{2}|\vec{\nabla} \phi|^{2}+f_{A B}(\phi, c, T)\right]. \label{F_functional}
\end{equation}
Inside the integrand, the first term is the gradient energy that ensures a finite interface thickness. The second term $f_{A B}(\phi, c, T)$ is the bulk free-energy density of a binary mixture of A and B atoms/molecules, where $c$ denotes the solute concentration defined as the mole fraction of B. In the dilute limit, the free-energy density can be written as
\begin{equation}
f_{A B}(\phi, c, T)=f\left(\phi, T_{M}\right)+f_{T}(\phi) (T-T_M)+\frac{R T_{M}}{v_{0}}(c \ln c-c)+\epsilon(\phi) c, \label{f_AB}
\end{equation}
where $f\left(\phi, T_{M}\right)$ is a double-well potential providing the stability of the two phases $\phi=\pm1$, and we choose a standard form
\begin{equation}
f\left(\phi, T_M\right) = h\left(-\frac{\phi^{2}}{2}+\frac{\phi^{4}}{4}\right),
\end{equation}
where $h$ denotes the energy barrier height between local minima corresponding to the solid and liquid. In Eq.\ \eqref{f_AB}, the free-energy density is written as the sum of the contribution due to solute addition and the contribution of the pure material, denoted by $f(\phi, T)$, which has been expanded to the first order in $(T-T_M)$ and $f_{T}(\phi) \equiv \partial f(\phi, T) /\partial T |_{T=T_{M}}$. The term $R T_M v_{0}^{-1}(c \ln c-c)$ is the standard entropy of mixing, where $R$ is the gas constant and $v_0$ is the molar volume assumed constant. The term $\epsilon(\phi) c$ is the enthalpy of mixing, where
\begin{equation}
\epsilon(\phi)=\frac{1+g(\phi)}{2} \epsilon_{s}+\frac{1-g(\phi)}{2} \epsilon_{l}    
\end{equation}
interpolates between the values $\epsilon_s$ and $\epsilon_l<\epsilon_s$ in the solid and liquid, respectively. Thus, we can also write it as $\epsilon(\phi)=[\bar{\epsilon}+g(\phi)\Delta \epsilon/2]$, where $\bar{\epsilon} \equiv(\epsilon_{s}+\epsilon_{l}) / 2$ and $\Delta \epsilon=(\epsilon_{s}-\epsilon_{l})$.

The system is expected to relax to a global free-energy minimum. The dynamical evolution of the scalar fields $c$ and $\phi$ follows standard variational forms for conserved and non-conserved dynamics, respectively:
\begin{align}
\frac{\partial c}{\partial t}&=\vec{\nabla} \cdot\left(K_{c} \vec{\nabla} \frac{\delta F}{\delta c}\right), \label{functional_dev_c} \\
\frac{\partial \phi}{\partial t}&=-K_{\phi} \frac{\delta F}{\delta \phi}, \label{functional_dev_p}
\end{align}
where $K_{c}$ is the mobility of solute atoms or molecules, and $K_{\phi}$ is a coefficient related to the interface kinetics. In equilibrium, $\partial_t \phi=\partial_t c=0$, and Eqs.\ \eqref{functional_dev_c}-\eqref{functional_dev_p} reduce to ${\delta F}/{\delta c}=\mu_{E}^{c}$ and ${\delta F}/{\delta \phi}=0$, where $\mu_{E}^{c}$ is the spatially uniform equilibrium value of the chemical potential. Applying the first equilibrium condition, one can obtain the stationary concentration profile $c_0(x)=c_l^0\exp(b\left[1+g(\phi_0(x))\right])$, where $\phi_0(x)$ is the stationary PF profile, $b =\ln k_e/2$, $k_e \equiv c_s^0/c_l^0 =\exp \left(-{v_{0} \Delta \epsilon}/{R T_{M}}\right)$, $c_s^0$ and $c_l^0$ are the equilibrium concentrations at the solid and liquid sides of the interface, respectively. Applying the second equilibrium condition, one can show that $\phi_0(x)$ is a tangent hyperbolic profile, and obtain the function $f_{T}(\phi)=(-{R T_{M}}/{v_{0} m_e}) \exp (b[1+g(\phi)])$ that reproduces the relation $T=(T_M-m_e c_l^0)$. To complete the model, we choose $K_c=v_{0} D_l q(\phi) c /\left(R T_{M}\right)$ such that Eq.\ \eqref{functional_dev_c} reproduces Fick's law of diffusion in the bulk phase. As a result, we obtain the evolution equations
\begin{align}
\tau_0 \frac{\partial \phi}{\partial t}=&W^{2} \nabla^{2} \phi + \phi -\phi^2 -\lambda g^{\prime}(\phi)\left[c+\frac{(T-T_M)}{m_e} \exp(b[1+g(\phi)])\right], \\
\frac{\partial c}{\partial t}=& \vec{\nabla} \cdot \left\{D_l q(\phi) c \vec{\nabla}[\ln c-b g(\phi)] \right\}, \label{eq2}
\end{align}
where we have defined $\tau_0=1 /\left(K_{\phi} h\right)$, $W=(\sigma / h)^{1 / 2}$, and $\lambda \equiv-b R T_{M} /\left(v_{0} h\right)>0$.

\section{Continuous growth model and asymptotic analyses}

In this supplemental section, we first introduce the solute trapping and solute drag in the continuous growth (CG) model \cite{Aziz1982ModelSolidification,Aziz1988ContinuousSolidification,Aziz1994OnSolidification}. Then, we derive the coefficients of the CG model, including a diffusive velocity $V_d$ and a solute drag coefficient $\alpha$, in the large-velocity asymptotic limit of the PF model with an interface thickness $W=W_0$, and interpolation functions $q(\phi)=(1-\phi)/2$ and $g(\phi)=15( \phi- 2\phi^{3}/3+\phi^{5}/5 )/8$.


In the CG model for ideal dilute binary alloys \cite{Aziz1994OnSolidification}, the non-equilibrium partition coefficient has the form
\begin{equation}
k(V)=\frac{k_e+V/V_d}{1+V/V_d}, \label{CGM_k}
\end{equation}
where $V$ is the interface velocity, and $V_d$ is the so-called diffusive velocity that generally depends on alloys. According to Eq.\ \eqref{CGM_k}, the non-equilibrium partition coefficient varies from $k(V) \to k_e$ at $V \ll V_d$ toward $k(V) \to 1$ at $V \gg V_d$. In the large-velocity limit, $V \gg V_d$, one can expand Eq.\ \eqref{CGM_k} to
\begin{equation}
k(V) \approx 1-\left(1-k_{e}\right)\left[\frac{V_{d}}{V}\right]+\mathcal{O} \left(\left[\frac{V_{d}}{V}\right]^{2}\right). \label{k_cg_approx}
\end{equation}
Meanwhile, the non-equilibrium liquidus slope in the dilute solution limit of the CG model has the form
\begin{equation}
\frac{m(V)}{m_e}=\frac{1 - k(V) + [k(V) + (1 - k(V))\alpha]\ln[k(V)/k_e] }{1-k_e}, \label{m_me}
\end{equation}
where $k(V)$ follows Eq.\ \eqref{CGM_k}. The coefficient $\alpha$ equals 1 for full solute drag and 0 if solute drag is negligible. According to Eq.\ \eqref{m_me}, the non-equilibrium liquidus slope varies from $m(V) \to m_e$ at $V \ll V_d$ toward $m(V) \to m_e{ \ln (1 / k_{e})}/{(1-k_e)}$ at $V \gg V_d$. One can expand Eq.\ \eqref{m_me} in the large-velocity limit
\begin{equation}
\frac{m(V)}{m_e} \approx \frac{\ln 1/k_e}{1-k_e}+\left(\alpha-1 \right) \ln 1/k_e \left[\frac{V_{d}}{V}\right]+\mathcal{O} \left(\left[\frac{V_{d}}{V}\right]^{2}\right). \label{m_v_limit}    
\end{equation}


\begin{figure}[hbt!]
\includegraphics[scale=1.2]{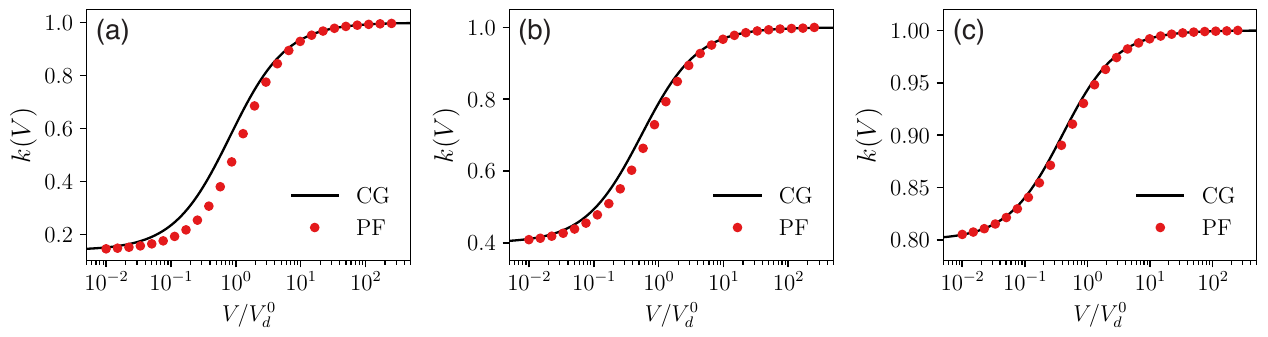}
\caption{\label{fig:figure_k} 
Comparison of $k(V)$ in the continuous growth (CG) model and the approximate solution of the PF model for (a) $k_e=0.14$, (b) $k_e=0.4$, and (c) $k_e=0.8$.
}
\end{figure}

A relation between the diffusive velocity $V_d$ in the CG model and the quantity $V_d^0 \equiv D_l/W_0$ defined in the PF model can be derived in the large-velocity asymptotic limit using the known method in Ref.\ \cite{Ahmad1998SoluteSolidification}. In the limit of $V \gg V_d^0$, we solve Eq.\ (5) of the main text and obtain
\begin{equation}
c(\phi) \approx c_{\infty}\left[1- \frac{V_{d}^0}{V} \frac{\ln1/k_e}{2} q(\phi) \frac{d g(\phi)}{d x}\right], \label{solution1}
\end{equation}
where $x$ is scaled by the interface thickness $W$. The 1D stationary solution of the PF has a standard tangent hyperbolic profile $\phi_0(x)=-\tanh{ ( {x}/{\sqrt{2}} ) }$, and the derivative of this stationary solution is $\partial_x \phi_0=-(1-\phi_0^2)/\sqrt{2}$. With the assumption that the PF profile for a moving interface remains close to its stationary value, the concentration profile is solely determined by Eq.\ \eqref{solution1}. Then, we substitute $\phi(x)=\phi_0(x)$ into Eq.\ \eqref{solution1} and obtain
\begin{equation}
c(\phi_0) \approx c_{\infty}\left[1+\frac{V_d^0}{V} \frac{15 \ln 1 / k_{e}}{32 \sqrt{2} }\left(1-\phi_{0}\right)^{4}\left(1+\phi_{0}\right)^{3} \right]. \label{c_phi0}
\end{equation}
The maximum value of $c(\phi_0)$ is found at $\phi_0=-1/7$. Thus, the partition coefficient is derived from Eq.\ \eqref{c_phi0} as $k=c_{\infty}/c_l=c_{\infty}/c(-1/7)$. Comparing this solution of partition coefficient to Eq.\ \eqref{k_cg_approx}, it is not difficult to find the relation between $V_d$ and $V_d^0$, which has the form
\begin{equation}
V_d=\frac{207360 \sqrt{2} \ln 1 / k_{e}}{823543(1-k_e)} V_d^0 \approx 0.356 \frac{ \ln 1 / k_{e}}{(1-k_e)} V_d^0. \label{V_d_limit}
\end{equation}
For $k_e=0.14$, $V_d \approx 0.814 V_d^0$.

In Fig.\ \ref{fig:figure_k}, we compare $k(V)$ in the CG model and the ``approximate solution'' of the PF model for different $k_e$. In the CG model, $V_d$ is calculated using Eq.\ \eqref{V_d_limit} with an input of $k_e$. The approximate solution is obtained by solving numerically Eq.\ (5) of the main text with the approximation that $\phi(x)=\phi_0(x)$. For $k_e=0.14$, the approximate solution of the PF model agrees well with the CG model at small ($V \ll V_d^0$) and large ($V \gg V_d^0$) velocities, but the deviation is found at intermediate velocities. However, as $k_e$ increases from 0.14 to 0.8, an almost perfect agreement is found over the entire velocity range.


Meanwhile, we use Eq.\ (8) of the main text to calculate the non-equilibrium liquidus slope in the large-velocity asymptotic limit, where both PF and concentration profiles are needed for evaluating the integral. As before, we assume that the PF profile for a moving interface remains close to its stationary value, $\phi(x)=\phi_0(x)$. For the concentration profile, we substitute Eq.\ \eqref{V_d_limit} into Eq.\ \eqref{c_phi0} and obtain
\begin{equation}
c(\phi_0) \approx c_{\infty}\left[1+\frac{V_d}{V} \frac{15 (1-k_e)}{32 \sqrt{2} C_1 }\left(1-\phi_{0}\right)^{4}\left(1+\phi_{0}\right)^{3}\right],
\end{equation}
where $C_1={207360 \sqrt{2}}/{823543}$. Then, according to Eq.\ (8) of the main text and the relation $k=c_{\infty}/c_{l} \approx 1-\left(1-k_{e}\right) V_{d} / V$, we obtain in the limit of $V \gg V_d$ that
\begin{equation}
\frac{m(V)}{m_e}\approx \frac{\ln 1/k_e}{1-k_e}+\left(\frac{25}{77 \sqrt{2}C_1}-1 \right) \ln 1/k_e \left[\frac{V_{d}}{V}\right]+\mathcal{O} \left(\left[\frac{V_{d}}{V}\right]^{2}\right). \label{m_v_m_e}
\end{equation}
Comparing Eq.\ \eqref{m_v_m_e} to Eq.\ \eqref{m_me}, it is not difficult to find the solute drag coefficient $\alpha=25/(77\sqrt{2}C_1)\approx0.645$, which is a constant for all $k_e$.

\begin{figure}[hbt!]
\includegraphics[scale=1.2]{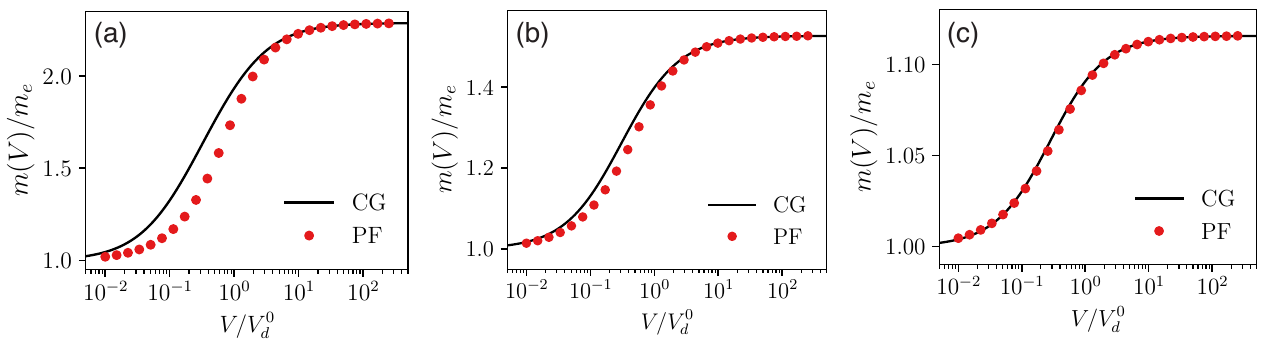}
\caption{\label{fig:figure_m} 
Comparison of $m(V)$ in the continuous growth (CG) model and the approximate solution of the PF model for (a) $k_e=0.14$, (b) $k_e=0.4$, and (c) $k_e=0.8$.
}
\end{figure}

In Fig.\ \ref{fig:figure_m}, we compare $m(V)$ in the CG model and the approximate solution of the PF model for different $k_e$. In the CG model, $V_d$ is calculated using Eq.\ \eqref{V_d_limit}, and $\alpha$ is a constant 0.645 according to the asymptotic analyses. $m(V)$ in the PF model is calculated using Eq.\ (8) of the main text with $\phi(x)=\phi_0(x)$ and the approximate solution of the concentration field (solution of Eq.\ (5) of the main text with $\phi(x)=\phi_0(x)$) as inputs. For $k_e=0.14$, the approximate solution of the PF model agrees well with the CG model at small ($V \ll V_d^0$) and large ($V \gg V_d^0$) velocities, but the deviation is found at intermediate velocities. As $k_e$ increases from 0.14 to 0.8, an almost perfect agreement is found over the entire velocity range. 

\section{Tip supersaturation}

The supersaturation $\Omega$ is defined as
\begin{equation}
\Omega=\frac{c_{l}-c_{\infty}}{c_{l}-c_{s}},
\end{equation}
where $c_l$ and $c_s$ measure solute concentrations at a given temperature at the liquid and solid sides of the interface, respectively. For a steady-state dendrite, the tip temperature reaches equilibrium, and we can obtain the tip supersaturation using the relation 
\begin{equation}
\Omega=\frac{T_{M}-T-m c_{\infty}-V / \mu_{k}}{(1-k)(T_{M}-T-{V}/{\mu_{k}})}, \label{omega}
\end{equation}
where the capillarity is not taken into account, and $\mu_k=(1+\epsilon_k)\mu_k^0$ in the $\left<10\right>$ dendrite growth direction. The non-equilibrium partition coefficient $k$ and liquidus slope $m$ in Eq.\ \eqref{omega} are obtained from the 1D full solution of the PF model at the same $V$. 

\begin{figure}[hbt!]
\includegraphics[scale=0.8]{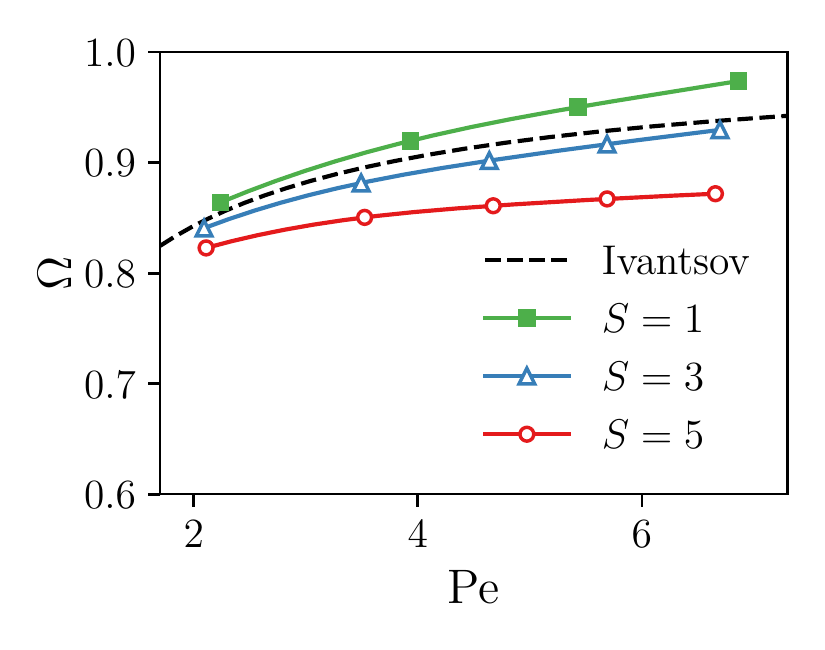}
\caption{\label{fig:figure_omega} 
Comparison of Ivantsov's solution and the tip supersaturation $\Omega$ measured in 2D PF simulations (with different interface thicknesses $S$) of dendritic array growth for Al-3wt.\% Cu and $G=5\times 10^6$ K/m; the primary spacing $\Lambda = 0.65$ $\mathrm{\mu m}$ and the simulated physical time is $10^{-4}$ s.
}
\end{figure}

The measured $\Omega$ in PF simulations are compared to the 2D Ivantsov's solution \cite{Ivantsov1947TemperatureMelt} 
\begin{equation}
\Omega(\mathrm{Pe})=\sqrt{\pi \, \mathrm{Pe}} \exp (\mathrm{Pe}) \mathrm{erfc}(\sqrt{\mathrm{Pe}}),    
\end{equation}
where $\mathrm{Pe}=R_{\mathrm{tip}} V/2D_l$ is the P\'eclet number. Note that Ivantsov's solution is obtained for isothermal dendrite growth by neglecting the capillarity and interface kinetics. As shown in Fig.\ \ref{fig:figure_omega}, the tip supersaturation $\Omega$ measured using Eq.\ \eqref{omega} in 2D PF simulations for directional solidification of an Al-3wt.\% Cu alloy are compared to Ivantsov's solution. Good agreements between simulations with different interface thicknesses and Ivantsov's solution are found at small Pe. For $W=W_0$ ($S=1$), the deviation between PF simulation and Ivantsov's solution slightly increases at a larger Pe, which indicates that the kinetics not included in Ivantsov's solution becomes more significant at a larger Pe, i.e., a larger tip velocity.


\bibliography{references}